\documentclass[aps,prd,reprint,twoside,superscriptaddress,showpacs,floatfix]{revtex4-1}

\usepackage{graphicx}
\usepackage{dcolumn}
\usepackage{bm}
\usepackage{natbib}
\usepackage{multirow}
\usepackage{amsmath}
\usepackage{amssymb}
\usepackage{mathrsfs}
\usepackage{bbold}
\usepackage{hyperref}
\usepackage{aas_macros}

\newcommand{\hMsun}{h^{-1}\mathrm{M}_{\odot}}
\newcommand{\dhalo}{\boldsymbol{\delta}_{\mathrm{h}}}
\newcommand{\dm}{\delta}
\newcommand{\dhh}{\delta_{\mathrm{h}}}
\newcommand{\dd}{\boldsymbol{\delta}}
\newcommand{\e}{\boldsymbol{\epsilon}}
\newcommand{\E}{\boldsymbol{\mathcal{E}}}
\newcommand{\Z}{\boldsymbol{\mathcal{A}}}
\newcommand{\C}{\mathbf{C}_{\mathrm{h}}}
\newcommand{\Cm}{\mathbf{C}}
\newcommand{\I}{\mathbf{I}}
\newcommand{\fnl}{f_{\mathrm{NL}}}
\newcommand{\bg}{\boldsymbol{b}}
\newcommand{\M}{\boldsymbol{M}}
\newcommand{\Mr}{\boldsymbol{\mathcal{M}}}
\newcommand{\np}{\boldsymbol{n}_\mathrm{P}}
\newcommand{\nc}{\boldsymbol{n}_\mathrm{c}}
\newcommand{\Vpm}{\boldsymbol{V}_{\!\!\pm}}
\newcommand{\Vmp}{\boldsymbol{V}_{\!\!\mp}}
\newcommand{\Vp}{\boldsymbol{V}_{\!\!+}}
\newcommand{\Vm}{\boldsymbol{V}_{\!\!-}}
\newcommand{\Om}{\Omega_\mathrm{m}}
\newcommand{\Ob}{\Omega_\mathrm{b}}
\newcommand{\rhom}{\bar{\rho}_\mathrm{m}}
\newcommand{\dc}{\delta_\mathrm{c}}
\newcommand{\Dc}{\Delta_\mathrm{c}}
\newcommand{\ns}{n_\mathrm{s}}
\newcommand{\T}{^{\phantom{\intercal}}}
\newcommand{\conv}{\!*\!}

\begin{document}

\title{Optimal Constraints on Local Primordial Non-Gaussianity from the Two-Point Statistics of Large-Scale Structure}

\author{Nico Hamaus} \email{hamaus@physik.uzh.ch}
\affiliation{Institute for Theoretical Physics, University of Zurich, 8057 Zurich, Switzerland}
\author{Uro{\v s} Seljak}
\affiliation{Institute for Theoretical Physics, University of Zurich, 8057 Zurich, Switzerland}
\affiliation{Physics Department, Astronomy Department and Lawrence Berkeley National Laboratory, University of California, Berkeley, California 94720, USA}
\affiliation{Ewha University, Seoul 120-750, S. Korea}
\author{Vincent Desjacques}
\affiliation{Institute for Theoretical Physics, University of Zurich, 8057 Zurich, Switzerland}

\date{\today}

\begin{abstract}
One of the main signatures of primordial non-Gaussianity of the local type is a scale-dependent correction to the bias of large-scale structure tracers such as galaxies or clusters, whose amplitude depends on the bias of the tracers itself. The dominant source of noise in the power spectrum of the tracers is caused by sampling variance on large scales (where the non-Gaussian signal is strongest) and shot noise arising from their discrete nature. Recent work has argued that one can avoid sampling variance by comparing multiple tracers of different bias, and suppress shot noise by optimally weighting halos of different mass. Here we combine these ideas and investigate how well the signatures of non-Gaussian fluctuations in the primordial potential can be extracted from the two-point correlations of halos and dark matter. On the basis of large $N$-body simulations with local non-Gaussian initial conditions and their halo catalogs we perform a Fisher matrix analysis of the two-point statistics. Compared to the standard analysis, optimal weighting and multiple-tracer techniques applied to halos can yield up to 1 order of magnitude improvements in $\fnl$-constraints, even if the underlying dark matter density field is not known. In this case one needs to resolve all halos down to $10^{10}\hMsun$ at $z=0$, while with the dark matter this is already achieved at a mass threshold of $10^{12}\hMsun$. We compare our numerical results to the halo model and find satisfactory agreement. Forecasting the optimal $\fnl$-constraints that can be achieved with our methods when applied to existing and future survey data, we find that a survey of $50h^{-3}\mathrm{Gpc}^3$ volume resolving all halos down to $10^{11}\hMsun$ at $z=1$ will be able to obtain $\sigma_{\fnl}\sim1$ ($68\%$ cl), a factor of $\sim20$ improvement over the current limits. Decreasing the minimum mass of resolved halos, increasing the survey volume or obtaining the dark matter maps can further improve these limits, potentially reaching the level of $\sigma_{\fnl}\sim0.1$. This precision opens up the possibility to distinguish different types of primordial non-Gaussianity and to probe inflationary physics of the very early Universe.
\end{abstract}


\maketitle

\setcounter{footnote}{0}

\section{Introduction}
A detection of primordial non-Gaussianity has the potential to test today's standard inflationary paradigm and its alternatives for the physics of the early Universe. Measurements of the CMB bispectrum furnish a direct probe of the nature of the initial conditions (see, e.g., \cite{Komatsu2010,Bartolo2010b,Liguori2010,Yadav2010,Komatsu2011} and references therein), but are limited by the two-dimensional nature of the CMB and its damping on small scales. However, the non-Gaussian signatures imprinted in the initial fluctuations of the potential gravitationally evolve into the large-scale structure (LSS) of the Universe, which can be observed in all three dimensions and whose statistical properties can be constrained with galaxy clustering data (for recent reviews, see \cite{Desjacques2010a,Verde2010}).

One of the cleanest probes is the galaxy (or, more generally, any tracer of LSS including clusters, etc.) two-point correlation function (in configuration space) or power spectrum (in Fourier space), which develops a characteristic scale dependence on large scales in the presence of primordial non-Gaussianity of the local type \cite{Dalal2008}. The power spectrum picks up an additional term proportional to $\fnl(b_{\mathrm{G}}-1)$, where $b_{\mathrm{G}}$ is the Gaussian bias of the tracer and $\fnl$ is a parameter describing the strength of the non-Gaussian signal. However, the precision to which we can constrain $\fnl$ is limited by sampling variance on large scales: each Fourier mode is an independent realization of a (nearly) Gaussian random field, so the ability to determine its rms-amplitude from a finite number of modes is limited. Recent work has demonstrated that it is possible to circumvent sampling variance by comparing two different tracers of the same underlying density field \cite{Seljak2009a,Slosar2009,McDonald2009a,Gil-Marin2010}. The idea is to take the ratio of power spectra from two tracers to (at least partly) cancel out the random fluctuations, leaving just the signature of primordial non-Gaussianity itself.

Another important limitation arises from the fact that galaxies are discrete tracers of the underlying dark matter distribution. Therefore, with a finite number of observable objects, the measurement of their power spectrum is affected by shot noise. Assuming galaxies are sampled from a Poisson process, this adds a constant contribution to their power spectrum, it is given by the inverse tracer number density $1/\bar{n}$. This is particularly important for massive tracers such as clusters, since their number density is very low. Yet they are strongly biased and therefore very sensitive to a potential non-Gaussian signal. Recent work has demonstrated the Poisson shot noise model to be inadequate \cite{Seljak2009b,Hamaus2010,Cai2011}. In particular, \cite{Seljak2009b,Hamaus2010} have shown that a mass-dependent weighting can considerably suppress the stochasticity between halos and the dark matter and thus reduce the shot noise contribution. In view of constraining primordial non-Gaussianity from LSS, this can be a very helpful tool to further reduce the error on $\fnl$.

Both of these methods (sampling variance cancellation and shot noise suppression) have so far been discussed separately in the literature. In this paper we combine the two to derive optimal constraints on $\fnl$ that can be achieved from two-point correlations of LSS. We show that dramatic improvements are feasible, but we do not imply that two-point correlations achieve optimal constraints in general: further gains may be possible when considering higher-order correlations, starting with the bispectrum analysis \cite{Baldauf2011a} (three-point correlations).

This paper is organized as follows: Sec.~\ref{sec:localng} briefly reviews the impact of local primordial non-Gaussianity on the halo bias, and the calculation of the Fisher information content on $\fnl$ from two-point statistics in Fourier space is presented in Sec.~\ref{sec:model}. In Sec.~\ref{sec:sims} we apply our weighting and multitracer methods to dark matter halos extracted from a series of large cosmological $N$-body simulations and demonstrate how we can improve the $\fnl$-constraints. These results are confronted with the halo model predictions in Sec.~\ref{sec:HM} before we finally summarize our findings in Sec.~\ref{sec:conclusion}.

\section{Non-Gaussian Halo Bias}
\label{sec:localng}

Primordial non-Gaussianity of the local type is usually characterized by expanding Bardeen's gauge-invariant potential $\Phi$ about the fiducial Gaussian case. Up to second order, it can be parametrized by the mapping \cite{Salopek1990,Gangui1994,Komatsu2001,Bartolo2004}
\begin{equation}
\Phi({\bf x})=\Phi_{\mathrm{G}}({\bf x})+\fnl\Phi_{\mathrm{G}}^2({\bf x}) \;,
\end{equation}
where $\Phi_{\mathrm{G}}({\bf x})$ is an isotropic Gaussian random field and $\fnl$ a dimensionless phenomenological parameter. Ignoring smoothing (we will consider scales much larger than the Lagrangian size of a halo), the linear density perturbation $\delta_0$ is related to $\Phi$ through the Poisson equation in Fourier space,
\begin{equation}
\delta_0({\bf k},z)=\frac{2}{3}\frac{k^2T(k)D(z)c^2}{\Om H_0^2}\Phi({\bf k}) \;,
\end{equation}
where $T(k)$ is the matter transfer function and $D(z)$ is the linear growth rate normalized to $1+z$. Applying the peak-background split argument to the Gaussian piece of Bardeen's potential, one finds a scale-dependent correction to the linear halo bias \cite{Dalal2008,Matarrese2008,Slosar2008}:
\begin{equation}
b(k,\fnl)=b_{\mathrm{G}}+\fnl(b_{\mathrm{G}}-1)u(k,z) \;, \label{b(k,fnl)}
\end{equation}
where $b_{\mathrm{G}}$ is the scale-independent linear bias parameter of the corresponding Gaussian field ($\fnl=0$) and
\begin{equation}
u(k,z)\equiv\frac{3\dc\Om H_0^2}{k^2T(k)D(z)c^2} \;. \label{u(k,z)}
\end{equation}
Here, $\dc\simeq1.686$ is the linear critical overdensity for spherical collapse. Corrections to Eq.~(\ref{b(k,fnl)}) beyond linear theory have already been worked out and agree reasonably well with numerical simulations \cite{Giannantonio2010,Jeong2009,Sefusatti2009b,McDonald2008}. Also, the dependence of the halo bias on merger history and halo formation time affects the amplitude of the non-Gaussian corrections in Eq.~(\ref{b(k,fnl)}) \cite{Slosar2008,Gao2005,Gao2007,Reid2010}, which we will neglect here.

\section{Fisher information from the two-point statistics of LSS \label{sec:model}}
It is believed that all discrete tracers of LSS, such as galaxies and clusters, reside within dark matter halos, collapsed nonlinear structures that satisfy the conditions for galaxy formation. The analysis of the full complexity of LSS is therefore reduced to the information content in dark matter halos. In this section we introduce our model for the halo covariance matrix and utilize it to compute the Fisher information content on $\fnl$ from the two-point statistics of halos and dark matter in Fourier space. We separately consider two cases: first halos only and second halos combined with dark matter. While the observation of halos is relatively easy with present-day galaxy redshift surveys, observing the underlying dark matter is hard, but not impossible: weak-lensing tomography is the leading candidate to achieve that.

\subsection{Covariance of Halos}

\subsubsection{Definitions}
We write the halo overdensity in Fourier space as a vector whose elements correspond to $N$ successive bins
\begin{equation}
\dhalo \equiv \left(\delta_\mathrm{h_1},\delta_\mathrm{h_2},\dots,\delta_{\mathrm{h}_N}\right)^\intercal \;.
\end{equation}
In this paper we will only consider a binning in halo mass, but the following equations remain valid for any quantity that the halo density field depends on (e.g., galaxy-luminosity, etc.). The covariance matrix of halos is defined as
\begin{equation}
\C\equiv\langle\dhalo\T\dhalo^\intercal\rangle \;,
\end{equation}
i.e., the outer product of the vector of halo fields averaged within a $k$-shell in Fourier space. Assuming the halos to be locally biased and stochastic tracers of the dark matter density field $\dm$, we can write
\begin{equation}
\dhalo = \bg\dm+\e \;, \label{dhalo}
\end{equation}
and we define
\begin{equation}
\bg\equiv\frac{\langle\dhalo\dm\rangle}{\langle\dm^2\rangle} \; \label{b}
\end{equation}
as the \emph{effective bias}, which is generally scale-dependent and non-Gaussian. $\e$ is a residual noise-field with zero mean and we assume it to be uncorrelated with the dark matter, i.e., $\langle\e\dm\rangle=0$ \cite{Manera2011}.

In each mass bin, the effective bias $\bg$ shows a distinct dependence on $\fnl$. In what follows, we will assume that $\bg$ is linear in $\fnl$, as suggested by Eq.~(\ref{b(k,fnl)}):
\begin{equation}
\bg(k,\fnl) = \bg_{\mathrm{G}}+\fnl\bg'(k) \;. \label{b_fnl}
\end{equation}
Here, $\bg_{\mathrm{G}}$ is the Gaussian effective bias and $\bg'\equiv\partial\bg/\partial\fnl$. Finally, we write $P\equiv\langle\dm^2\rangle$ for the nonlinear dark matter power spectrum and assume $\partial P/\partial\fnl=0$. This is a good approximation on large scales \cite{Desjacques2009,Pillepich2010,Smith2011}. Thus, the model from Eq.~(\ref{dhalo}) yields the following halo covariance matrix:
\begin{equation}
\C=\bg\bg^\intercal P + \E \;, \label{Cov}
\end{equation}
where the \emph{shot noise matrix} $\E$ was defined as
\begin{equation}
\E\equiv\langle\e\e^\intercal\rangle \;.
\end{equation}
In principle, $\E$ can contain other components than pure Poisson noise, for instance higher-order terms from the bias expansion \cite{Fry1993,Mo1996,McDonald2006}. Here and henceforth, we will define $\E$ as the residual from the effective bias term $\bg\bg^\intercal P$ in $\C$, and allow it to depend on $\fnl$. Thus, with Eqs.~(\ref{b}) and~(\ref{Cov}) the shot noise matrix can be written as
\begin{equation}
\E = \langle\dhalo\T\dhalo^\intercal\rangle - \frac{\langle\dhalo\T\dm\rangle\langle\dhalo^\intercal\dm\rangle}{\langle\dm^2\rangle} \;. \label{E}
\end{equation}
This agrees precisely with the definition given in \cite{Hamaus2010} for the Gaussian case, however it also takes into account the possibility of a scale-dependent effective bias in non-Gaussian scenarios, such that the effective bias term $\bg\bg^\intercal P$ always cancels in this expression \cite{Kendrick2010}.

Reference~\cite{Slosar2009} already investigated the Fisher information content on primordial non-Gaussianity for the idealized case of a purely Poissonian shot noise component in the halo covariance matrix. In \cite{Cai2011}, the halo covariance was suggested to be of a similar simple form, albeit with a modified definition of halo bias and a diagonal shot noise matrix. In this work we will consider the more general model of Eq.~(\ref{Cov}) without assuming anything about $\E$. Instead we will investigate the shot noise matrix with the help of $N$-body simulations.

The Gaussian case has already been studied in \cite{Hamaus2010}. Simulations revealed a very simple eigenstructure of the shot noise matrix: for $N>2$ mass bins of equal number density $\bar{n}$ it exhibits a $(N-2)$-dimensional degenerate subspace with eigenvalue $\lambda_{\mathrm{P}}^{\left(N-2\right)}=1/\bar{n}$, which is the expected result from Poisson sampling. Of the two remaining eigenvalues $\lambda_{\pm}$, one is enhanced ($\lambda_+$) and one suppressed ($\lambda_-$) with respect to the value $1/\bar{n}$. The shot noise matrix can thus be written as
\begin{equation}
\E=\bar{n}^{-1}\I+(\lambda_{+}-\bar{n}^{-1})\Vp\T\Vp^\intercal+(\lambda_{-}-\bar{n}^{-1})\Vm\T\Vm^\intercal \;, \label{E_eb}
\end{equation}
where $\I$ is the $N\times N$ identity matrix and $\Vpm$ are the normalized eigenvectors corresponding to $\lambda_{\pm}$. Its inverse takes a very similar form
\begin{equation}
\E^{-1}=\bar{n}\I+(\lambda_{+}^{-1}-\bar{n})\Vp\T\Vp^\intercal+(\lambda_{-}^{-1}-\bar{n})\Vm\T\Vm^\intercal \;.
\end{equation}
The halo model \cite{Seljak2000} can be applied to predict the functional form of $\lambda_{\pm}$ and $\Vpm$ (see \cite{Hamaus2010} and Sec.~\ref{sec:HM}). This approach is however not expected to be exact, as it does not ensure mass- and momentum conservation of the dark matter density field and leads to white-noise-like contributions in both the halo-matter cross and the matter auto power spectra which are not observed in simulations \cite{Crocce2008}. Yet, the halo model is able to reproduce the eigenstructure of $\E$ fairly well \cite{Hamaus2010} and we will use it for making predictions beyond our $N$-body resolution limit.

In the Gaussian case one can also relate the dominant eigenmode $\Vp$ with corresponding eigenvalue $\lambda_{+}$ to the second-order term arising in a local bias-expansion model \cite{Fry1993,Mo1996}, where the coefficients $\bg_i$ are determined analytically from the peak-background split formalism given a halo mass function \cite{Sheth1999,Scoccimarro2001}. In non-Gaussian scenarios this can be extended to a multivariate expansion in dark matter density $\dm$ and primordial potential $\Phi$ including bias coefficients for both fields \cite{Giannantonio2010,Baldauf2011a}. For the calculation of $\E$ we will however restrict ourselves to the Gaussian case and later compare with the numerical results of non-Gaussian initial conditions to see the effects of $\fnl$ on $\E$ and its eigenvalues.
The suppressed eigenmode $\Vm$ with eigenvalue $\lambda_{-}$ can also be explained by a halo-exclusion correction to the Poisson-sampling model for halos, as studied in \cite{Smith2011}.

In what follows, we will truncate the local bias expansion at second order. Therefore, we shall assume the following model for the halo overdensity in configuration space
\begin{equation}
\dhalo(\mathbf{x}) = \bg_1\dm(\mathbf{x}) + \bg_2\dm^2(\mathbf{x}) + \np(\mathbf{x}) + \nc(\mathbf{x}) \;.
\end{equation}
Here, $\np$ is the usual Poisson noise and $\nc$ a correction to account for deviations from the Poisson-sampling model. In Fourier space, this yields
\begin{equation}
\dhalo(\mathbf{k}) = \bg_1\dm(\mathbf{k}) + \bg_2\left(\dm\conv\dm\right)(\mathbf{k}) + \np(\mathbf{k}) + \nc(\mathbf{k}) \;, \label{model}
\end{equation}
where the asterisk-symbol denotes a convolution. The Poisson noise $\np$ arises from a discrete sampling of the field $\dhalo$ with a finite number of halos, it is uncorrelated with the underlying dark matter density, $\langle\np\dm\rangle=0$, and its power spectrum is $\langle\np\T\np^\intercal\rangle=1/\bar{n}$ (Poisson white noise). We further assume $\langle\np\nc^\intercal\rangle=\langle\nc\dm\rangle=0$, which leads to
\begin{equation}
\bg = \bg_1+\bg_2\frac{\langle\left(\dm\conv\dm\right)\dm\rangle}{\langle\dm^2\rangle} \;, \label{b_model}
\end{equation}
\begin{align}
\C &= \bg_1\T\bg_1^\intercal\langle\dm^2\rangle + \left(\bg_1\T\bg_2^\intercal+\bg_2\T\bg_1^\intercal\right)\langle\left(\dm\conv\dm\right)\dm\rangle
\nonumber \\
&\quad +\bg_2\T\bg_2^\intercal\langle\left(\dm\conv\dm\right)^2\rangle + \langle\np\T\np^\intercal\rangle + \langle\nc\T\nc^\intercal\rangle\;,
\end{align}
\begin{equation}
\E = \bar{n}^{-1}\I + \bg_2\T\bg_2^\intercal\left[\langle\left(\dm\conv\dm\right)^2\rangle-\frac{\langle\left(\dm\conv\dm\right)\dm\rangle^2}{\langle\dm^2\rangle}\right] + \langle\nc\T\nc^\intercal\rangle \; . \label{E-model}
\end{equation}
Hence, we can identify the normalized vector $\bg_2/|\bg_2|$ with the eigenvector $\Vp$ of Eq.~(\ref{E_eb}) with corresponding eigenvalue
\begin{equation}
\lambda_+ = \bg_2^\intercal\bg_2\T\mathcal{E}_{\dm^2} + \bar{n}^{-1} \; , \label{lambda-model}
\end{equation}
where we define
\begin{equation}
\mathcal{E}_{\dm^2} \equiv \langle\left(\dm\conv\dm\right)^2\rangle-\frac{\langle\left(\dm\conv\dm\right)\dm\rangle^2}{\langle\dm^2\rangle} \; . \label{sigma_dm2}
\end{equation}

In \cite{McDonald2006} this term is absorbed into an effective shot noise power, since it behaves like white noise on large scales and arises from the peaks and troughs in the dark matter density field being nonlinearly biased by the $b_2$-term \cite{Heavens1998}. We evaluated $\mathcal{E}_{\dm^2}$ along with the expressions that appear in Eq.~(\ref{sigma_dm2}) with the help of our dark matter $N$-body simulations for Gaussian and non-Gaussian initial conditions (for details about the simulations, see Sec.~\ref{sec:sims}).

\begin{figure}[!t]
\centering
\resizebox{\hsize}{!}{
\includegraphics[trim=8 0 0 0,clip]{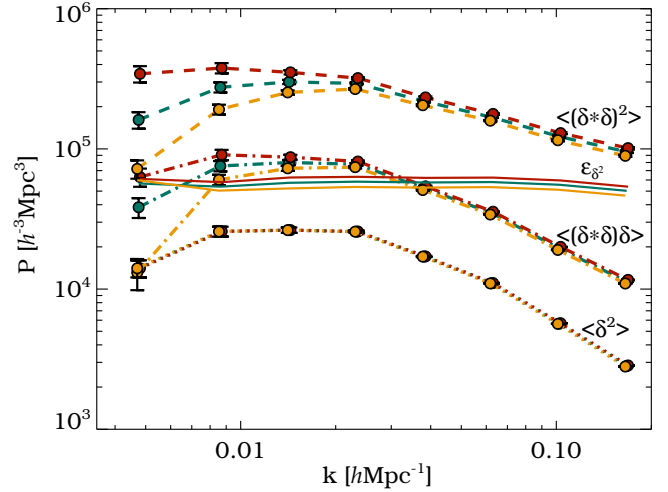}}
\caption{Shot noise $\mathcal{E}_{\dm^2}$ of the squared dark matter density field $\dm^2$ as defined in Eq.~(\ref{sigma_dm2}) with both Gaussian (solid green) and non-Gaussian initial conditions with $\fnl=+100$ (solid red) and $\fnl=-100$ (solid yellow) from $N$-body simulations at $z=0$. Clearly, $\mathcal{E}_{\dm^2}$ is close to white-noise like in all three cases. The auto power spectrum $\langle\left(\dm\conv\dm\right)^2\rangle$ of $\dm^2$ in Fourier space (dashed), its cross power spectrum $\langle\left(\dm\conv\dm\right)\dm\rangle$ with the ordinary dark matter field $\dm$ (dot-dashed), as well as the ordinary dark matter power spectrum $\langle\dm^2\rangle$ (dotted) are overplotted for the corresponding values of $\fnl$. The squared dark matter field $\dm^2$ can be interpreted as a biased tracer of $\dm$ and therefore shows the characteristic $\fnl$-dependence of biased fields (like halos) on large scales.}
\label{sn_m2}
\end{figure}

The results are depicted in Fig.~\ref{sn_m2}. $\mathcal{E}_{\dm^2}$ obviously shows a slight dependence on $\fnl$, but it remains white-noise-like even in the non-Gaussian cases. The $\fnl$-dependence of this term has not been discussed in the literature
yet, but it can have a significant impact on the power spectrum of high-mass halos which have a large $b_2$-term; see Eq.~(\ref{lambda-model}). A discussion of the numerical results for halos, specifically the $\fnl$-dependence of $\lambda_+$, is conducted later in this paper. It is also worth noticing the $\fnl$-dependence of $\langle\left(\dm\conv\dm\right)^2\rangle$ and $\langle\left(\dm\conv\dm\right)\dm\rangle$. The properties of the squared dark matter field $\dm^2(\mathbf{x})$ are similar to the ones of halos, namely, the $k^{-2}$-correction of the effective bias in Fourier space, which in this case is defined as $b_{\dm^2}\equiv\langle\left(\dm\conv\dm\right)\dm\rangle/\langle\dm^2\rangle$ and appears in Eq.~(\ref{b_model}).

The last term in Eq.~(\ref{E-model}) corresponds to the suppressed eigenmode of the shot noise matrix. Both its eigenvector and eigenvalue can be described reasonably well by the halo model \cite{Hamaus2010}. The argument of \cite{Smith2011} based on halo exclusion yields a similar result while providing a more intuitive explanation for the occurrence of such a term.

\subsubsection{Likelihood and Fisher information}
In order to find the \emph{best unbiased estimator} for $\fnl$, we have to maximize the likelihood function. Although we are dealing with non-Gaussian statistics of the density field, deviations from the Gaussian case are usually small in practical applications (e.g., \cite{Slosar2008,Carbone2008,Afshordi2008}), so we will consider a multivariate Gaussian likelihood
\begin{equation}
\mathscr{L}=\frac{1}{(2\pi)^{N/2}\sqrt{\det\C}}\exp\left(-\frac{1}{2}\dhalo^\intercal\C^{-1}\dhalo\T\right) \;. \label{likelihood}
\end{equation}
Maximizing this likelihood function is equivalent to minimizing the following chi-square,
\begin{equation}
\chi^2=\dhalo^\intercal\C^{-1}\dhalo\T+\ln\left(1+\alpha\right)+\ln(\det\E) \;, \label{chi2}
\end{equation}
where we dropped the irrelevant constant $N\ln(2\pi)$ and used
\begin{equation}
\det\C=\det\left(\bg\bg^\intercal P+\E\right)=(1+\alpha)\det\E \;,
\end{equation}
with $\alpha\equiv\bg^\intercal\E^{-1}\bg P$. For a single mass bin, Eq.~(\ref{chi2}) simplifies to
\begin{equation}
\chi^2=\frac{\dhh^2}{b^2P+\mathcal{E}}+\ln\left(b^2P+\mathcal{E}\right) \;. \label{chi2_1}
\end{equation}

The Fisher information matrix \cite{Fisher1935} for the parameters $\theta_i$ and $\theta_j$ and the random variable $\dhalo$ with covariance~$\C$, as derived from a multivariate Gaussian likelihood \cite{Tegmark1997,Heavens2009}, reads
\begin{equation}
F_{ij} \equiv \frac{1}{2}\mathrm{Tr}\left(\frac{\partial\C}{\partial\theta_i}\C^{-1}\frac{\partial\C}{\partial\theta_j}\C^{-1}\right) \;. \label{fisher}
\end{equation}
With the above assumptions, the derivative of the halo covariance matrix with respect to the parameter $\fnl$ is
\begin{equation}
\frac{\partial\C}{\partial\fnl} = \left(\bg\bg'^\intercal+\bg'\bg^\intercal\right)P + \E' \;, \label{C_fnl}
\end{equation}
with $\E'\equiv \partial\E/\partial\fnl$.
The inverse of the covariance matrix can be obtained by applying the \emph{Sherman-Morrison} formula \cite{Sherman1950,Bartlett1951}
\begin{equation}
\C^{-1}=\E^{-1}-\frac{\E^{-1}\bg\bg^\intercal\E^{-1}P}{1+\alpha} \;, \label{Sherman-Morrison}
\end{equation}
where again $\alpha\equiv\bg^\intercal\E^{-1}\bg P$. On inserting the two previous relations into Eq.~(\ref{fisher}), we eventually obtain the full expression for $F_{\fnl\fnl}$ in terms of $\bg$, $\bg'$, $\E$, $\E'$ and $P$ (see Appendix \ref{appendix1} for the derivation of Eq.~(\ref{F})). Neglecting the $\fnl$-dependence of $\E$, i.e., setting $\E'\equiv0$, the Fisher information on $\fnl$ becomes
\begin{equation}
F_{\fnl\fnl}=\frac{\alpha\gamma+\beta^2+\alpha\left(\alpha\gamma-\beta^2\right)}{\left(1+\alpha\right)^2} \;, \label{F_lin}
\end{equation}
with $\alpha\equiv\bg^\intercal\E^{-1}\bg P$, $\beta\equiv\bg^\intercal\E^{-1}\bg'P$ and $\gamma\equiv\bg'^\intercal\E^{-1}\bg'P$.
For a single mass bin, Eq.~(\ref{F}) simplifies to Eq.~(\ref{F1}),
\begin{equation}
F_{\fnl\fnl}=2\left(\frac{bb'P+\mathcal{E}'/2}{b^2P+\mathcal{E}}\right)^2 \;. \label{F1_text}
\end{equation}
This implies that even in the limit of a very well-sampled halo density field ($\bar{n}\rightarrow\infty$) with negligible shot noise power $\mathcal{E}$ (and neglecting $\mathcal{E}'$) the Fisher information content on $\fnl$ that can be extracted per mode from a single halo mass bin is limited to the value $2\left(b'/b\right)^2$. This is due to the fact that we can only constrain $\fnl$ from a change in the halo bias relative to the Gaussian expectation, not from a measurement of the effective bias itself. The latter can only be measured directly if one knows the dark matter distribution, as will be shown in the subsequent paragraph. However, the situation changes for several halo mass bins (multiple tracers as in \cite{Seljak2009a}). In this case, the Fisher information content from Eqs.~(\ref{F}) and (\ref{F_lin}) can exceed the value $2\left(b'/b\right)^2$ (see Sec.~\ref{sec:sims} and \ref{sec:HM}).

\subsection{Covariance of Halos and Dark Matter}

\subsubsection{Definitions}
We will now assume that we possess knowledge about the dark matter distribution in addition to the halo density field. In practice one may be able to achieve this by combining galaxy redshift surveys with lensing tomography \cite{Pen2004}, but the prospects are somewhat uncertain. We will simply add the dark matter overdensity mode $\dm$ to the halo overdensity vector $\dhalo$, defining a new vector
\begin{equation}
\dd \equiv \left(\dm,\delta_\mathrm{h_1},\delta_\mathrm{h_2},\dots,\delta_{\mathrm{h}_N}\right)^\intercal \;.
\end{equation}
In analogy with the previous section, we define the covariance matrix as $\Cm\equiv\langle\dd\dd^\intercal\rangle$ and write
\begin{equation}
\Cm = \left( \begin{array}{cc}
\langle\dm^2\rangle & \langle\dhalo^\intercal\dm\rangle \vspace{1pt} \\
\langle\dhalo\dm\rangle & \langle\dhalo\T\dhalo^\intercal\rangle \\
\end{array} \right) =
\left( \begin{array}{cc}
P & \bg^\intercal P \\
\bg P & \C \\
\end{array} \right) \;. \label{Covm}
\end{equation}

\subsubsection{Likelihood and Fisher information}
Upon inserting the new covariance matrix into the Gaussian likelihood as defined in Eq.~(\ref{likelihood}), we find the chi-square to be
\begin{equation}
\chi^2=\dd^\intercal\Cm^{-1}\dd\T+\ln{\left(\det\E\right)} \;, \label{chi2_Cm}
\end{equation}
where we used
\begin{equation}
\det\Cm=\det\C\det\left(P-\bg^\intercal\C^{-1}\bg P^2\right)=P\det\E \;,
\end{equation}
and we still assume $P$ to be independent of $\fnl$ and therefore drop the term $\ln{(P)}$ in Eq.~(\ref{chi2_Cm}). In terms of the halo and dark matter overdensities, the chi-square can also be expressed as
\begin{equation}
\chi^2=\left(\dhalo-\bg\dm\right)^\intercal\E^{-1}\left(\dhalo-\bg\dm\right)+\ln{\left(\det\E\right)} \;, \label{chi2m}
\end{equation}
which is equivalent to the definition in \cite{Hamaus2010} (where the last term was neglected). The corresponding expression for a single halo mass bin reads
\begin{equation}
\chi^2=\frac{\left(\dhh-b\dm\right)^2}{\mathcal{E}}+\ln{\left(\mathcal{E}\right)} \; . \label{chi2m_1}
\end{equation}
For the derivative of $\Cm$ with respect to $\fnl$ we get
\begin{equation}
\frac{\partial\Cm}{\partial\fnl} = \left( \begin{array}{cc}
0 & \bg'^\intercal P \\
\bg'P\;\; &\;\; \bg\bg'^\intercal P+\bg'\bg^\intercal P +\E' \\
\end{array} \right) \;. \label{Covm_fnl}
\end{equation}
Performing a block inversion, we readily obtain the inverse covariance matrix,
\begin{equation}
\Cm^{-1} = \left( \begin{array}{cc}
(1+\alpha)P^{-1} & -\bg^\intercal\E^{-1} \\
-\E^{-1}\bg & \E^{-1} \\
\end{array} \right) \;. \label{CovmI}
\end{equation}
As shown in Appendix \ref{appendix2}, the Fisher information content on $\fnl$ now becomes
\begin{equation}
F_{\fnl\fnl}=\gamma+\tau \;, \label{F_m_text}
\end{equation}
with $\gamma\equiv\bg'^\intercal\E^{-1}\bg' P$ and $\tau\equiv\frac{1}{2}\mathrm{Tr}\left(\E'\E^{-1}\E'\E^{-1}\right)$. For a single halo mass bin this further simplifies to
\begin{equation}
F_{\fnl\fnl}=\frac{b'^2P}{\mathcal{E}}+\frac{1}{2}\left(\frac{\mathcal{E}'}{\mathcal{E}}\right)^2 \;. \label{F_m1}
\end{equation}
It is worth noting that, in contrast to Eq.~(\ref{F1_text}), the Fisher information from one halo mass bin with knowledge of the dark matter becomes infinite in the limit of vanishing $\mathcal{E}$. In this limit the effective bias can indeed be determined exactly, allowing an exact measurement of $\fnl$~\cite{Seljak2009a}.

\section{Application to N-body simulations}
\label{sec:sims}
We employ numerical $N$-body simulations with both Gaussian and non-Gaussian initial conditions to find signatures of primordial non-Gaussianity in the two-point statistics of the final density fields in Fourier space. More precisely, we consider an ensemble of $12$ realizations of box-size $1.6h^{-1}\mathrm{Gpc}$ (this yields a total effective volume of $V_{\mathrm{eff}}\simeq50h^{-3}\mathrm{Gpc}^3$). Each realization is seeded with both Gaussian ($\fnl=0$) and non-Gaussian ($\fnl=\pm100$) initial conditions of the local type \cite{Desjacques2009}, and evolves $1024^3$ particles of mass $3.0\times10^{11}\hMsun$. The cosmological parameters are $\Om=0.279$, $\Omega_{\Lambda}=0.721$, $\Ob=0.046$, $\sigma_8=0.81$, $\ns=0.96$, and $h=0.7$, consistent with the \scshape wmap5 \rm \cite{Komatsu2009} best-fit constraint. Additionally, we consider one realization with each $\fnl=0,\pm50$ of box-size $1.3h^{-1}\mathrm{Gpc}$ with $1536^3$ particles of mass $4.7\times10^{10}\hMsun$ to assess a higher-resolution regime. The simulations were performed on the supercomputer \scshape zbox3 \rm at the University of Z\" urich with the \scshape gadget ii \rm code \cite{Springel2005b}. The initial conditions were laid down at redshift $z=100$ by perturbing a uniform mesh of particles with the Zel'dovich approximation.

To generate halo catalogs, we employ a friends-of-friends (FOF) algorithm \cite{Davis1985} with a linking length equal to $20\%$ of the mean interparticle distance. For comparison, we also generate halo catalogs using the \scshape ahf \rm halo finder developed by \cite{Gill2004}, which is based on the spherical overdensity (SO) method \cite{Lacey1994}. In this case, we assume an overdensity threshold $\Dc(z)$ being a decreasing function of redshift, as dictated by the solution to the spherical collapse of a tophat perturbation in a $\Lambda$CDM Universe \cite{Eke1996}. In both cases, we require a minimum of 20 particles per halo, which corresponds to a minimum halo mass $M_{\mathrm{min}}\simeq 5.9\times10^{12}\hMsun$ for the simulations with $1024^3$ particles. For Gaussian initial conditions the resulting total number density of halos is $\bar{n}\simeq 7.0\times10^{-4}h^3\mathrm{Mpc^{-3}}$ and $4.2\times10^{-4}h^3\mathrm{Mpc^{-3}}$ for the FOF and SO catalogs, respectively. Note that the FOF mass estimate is on average $20\%$ higher than the SO mass estimate. For our $1536^3$-particles simulation we obtain $M_{\mathrm{min}}\simeq 9.4\times10^{11}\hMsun$ and $\bar{n}\simeq 4.0\times10^{-3}h^3\mathrm{Mpc^{-3}}$ resulting from the FOF halo finder.

The binning of the halo density field into $N$ consecutive mass bins is done by sorting all halos by increasing mass and dividing this ordered array into $N$ bins with an equal number of halos. The halos of each bin $i\in\left[1\dots N\right]$ are selected separately to construct the halo density field~$\delta_{\mathrm{h_i}}$. The density fields of dark matter and halos are first computed in configuration space via interpolation of the particles onto a cubical mesh with $512^3$ grid points using a cloud-in-cell mesh assignment algorithm \cite{Hockney1988}. We then perform a fast fourier transform to compute the modes of the fields in $k$-space.

For each of our Gaussian and non-Gaussian realizations, we match the total number of halos to the one realization with the least amount of them by discarding halos from the low-mass end. This \emph{abundance matching} technique ensures that we eliminate any possible signature of primordial non-Gaussianity induced by the unobservable $\fnl$-dependence of the halo mass function. It guarantees a constant value $1/\bar{n}$ of the Poisson noise for both Gaussian and non-Gaussian realizations. A dependence of the Poisson noise on $\fnl$ would complicate the interpretation of the Fisher information content. Note also that, in order to calculate the derivative of a function $\mathcal{F}$ with respect to $\fnl$, we apply the linear approximation
\begin{equation}
\frac{\partial\mathcal{F}}{\partial\fnl}\simeq\frac{\mathcal{F}(\fnl=+100)-\mathcal{F}(\fnl=-100)}{2\times100}\;, \label{df}
\end{equation}
which exploits the statistics of all our non-Gaussian runs. All the error bars quoted in this paper are computed from the variance amongst our $12$ realizations.

\subsection{Effective bias and shot noise}
At the two-point level and in Fourier space, the clustering of halos as described by Eq.~(\ref{Cov}) is determined by two basic components: effective bias and shot noise. Since the impact of primordial non-Gaussianity on the nonlinear dark matter power spectrum $P$ is negligible on large scales (see Fig.~\ref{sn_m2}), the dependence of both $\bg$ and $\E$ on $\fnl$ must be known if one wishes to constrain the latter. In the following sections, we will examine this dependence in our series of $N$-body simulations.

\begin{figure*}[!t]
\centering
\resizebox{\hsize}{!}{
\includegraphics{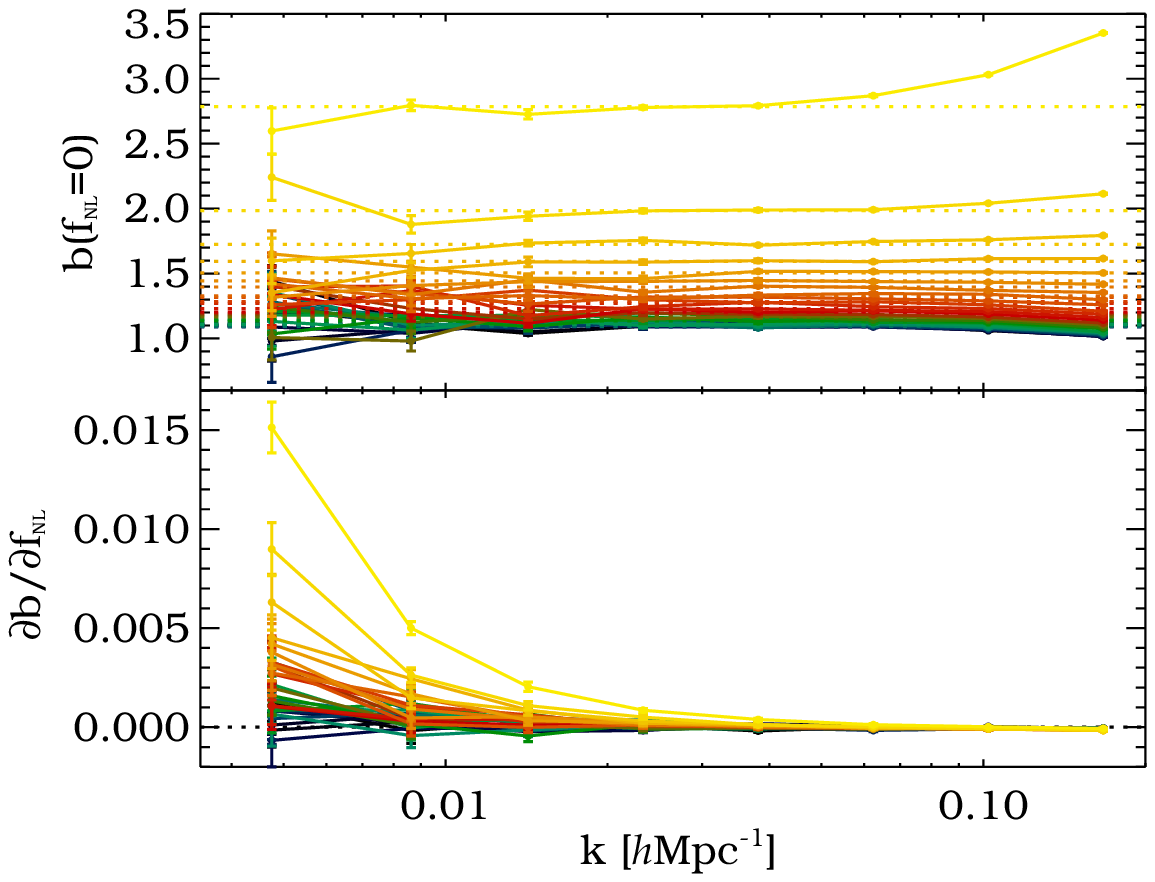}
\includegraphics{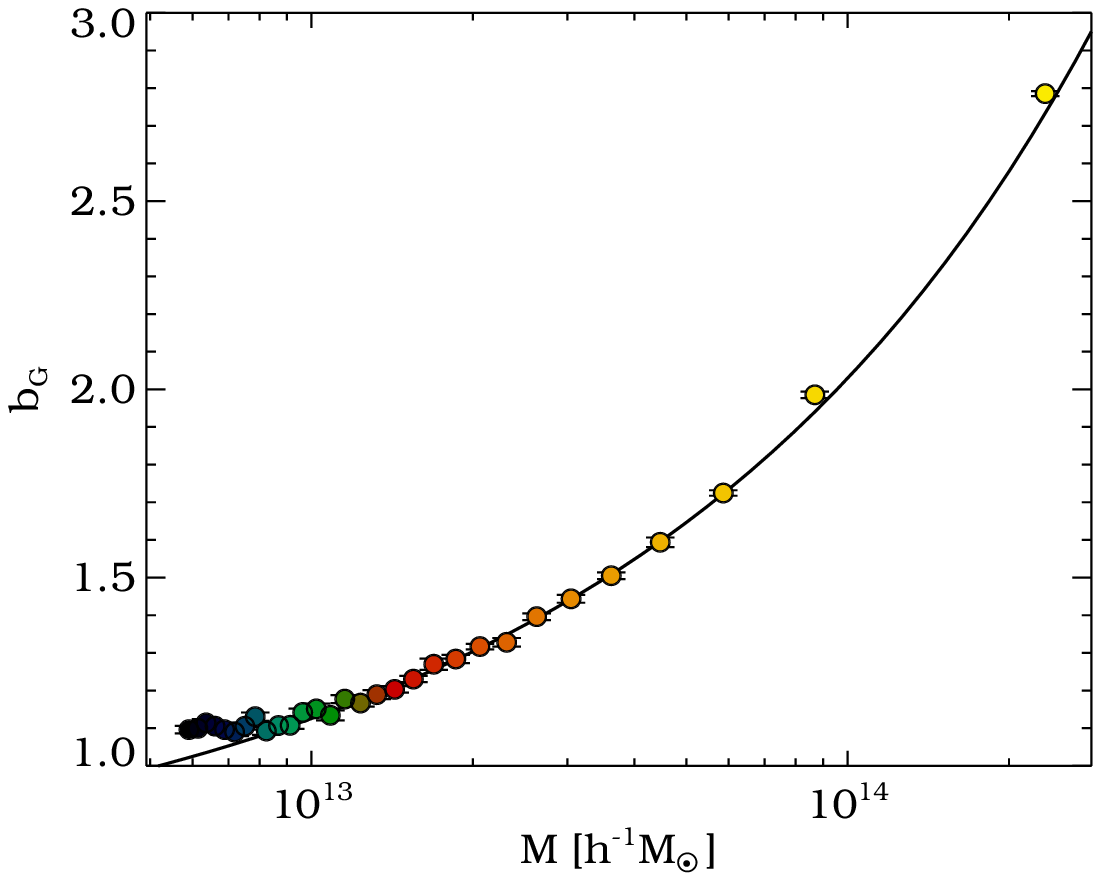}}
\caption{LEFT: Gaussian effective bias (top) and its derivative with respect to $\fnl$ (bottom) for the case of $30$ mass bins. The scale-independent part $\bg_{\mathrm{G}}$ is plotted in dotted lines for each bin; it was obtained by averaging all modes with $k\le0.032h\mathrm{Mpc}^{-1}$. RIGHT: Large-scale averaged Gaussian effective bias $\bg_{\mathrm{G}}$ from the left panel (dotted lines) plotted against mean halo mass. The solid line depicts the linear-order bias derived from the peak-background split formalism. All error bars are obtained from the variance of our $12$ boxes to their mean. Results are shown for FOF halos at $z=0$.}
\label{bias}
\end{figure*}

\subsubsection{Effective bias}
In the top left panel of Fig.~\ref{bias}, the effective bias $\bg$ in the fiducial Gaussian case ($\fnl=0$) is shown for 30 consecutive FOF halo mass bins as a function of wave number. In the large-scale limit $k\to 0$, the measurements are consistent with being scale-independent, as indicated by the dotted lines which show the average of $\bg(k,\fnl=0)$ over all modes with $k\le0.032h\mathrm{Mpc}^{-1}$, denoted $\bg_{\mathrm{G}}$. At larger wave numbers, the deviations can be attributed to higher-order bias terms, which are most important at high mass. Relative to the low-$k$ averaged, scale-independent Gaussian bias $\bg_{\mathrm{G}}$, these corrections tend to suppress the effective bias at low mass, whereas they increase it at the very high-mass end (see Eq.~(\ref{b_model})). The right panel of Fig.~\ref{bias} shows the large-scale average $\bg_{\mathrm{G}}$ as a function of halo mass, as determined from $30$ halo mass bins, each with a number density of $\bar{n}\simeq 2.3\times10^{-5}h^3\mathrm{Mpc^{-3}}$. The solid line is the linear-order bias as derived from the peak-background split formalism \cite{Sheth1999,Scoccimarro2001}. We find a good agreement with our $N$-body data, only at masses below $\sim8\times10^{12}\hMsun$ deviations appear for halos with less than $\sim30$ particles \cite{Knebe2011}.

The bottom left panel of Fig.~\ref{bias} depicts the derivative of $\bg$ with respect to $\fnl$ for each of the $30$ mass bins. The behavior is well described by the linear theory prediction of Eq.~(\ref{b(k,fnl)}), leading to a $k^{-2}$-dependence on large scales which is more pronounced for more massive halos (for quantitative comparisons with simulations, see \cite{Desjacques2009,Grossi2009,Pillepich2010}). Thus, the amplitude of this effect gradually diminishes towards smaller scales and even disappears around $k\sim0.1h\mathrm{Mpc}^{-1}$. Note that \cite{Desjacques2009} argued for an additional non-Gaussian bias correction which follows from the $\fnl$-dependence of the mass function. This $k$-independent contribution should in principle be included in Eq.~(\ref{b(k,fnl)}). However, as can be seen in the lower left plot, it is negligible in our approach (i.e., all curves approach zero at high $k$) owing to the matching of halo abundances between our Gaussian and non-Gaussian realizations.

\subsubsection{Shot noise matrix \label{shot noise matrix}}
\begin{figure*}[!t]
\centering
\resizebox{\hsize}{!}{
\includegraphics{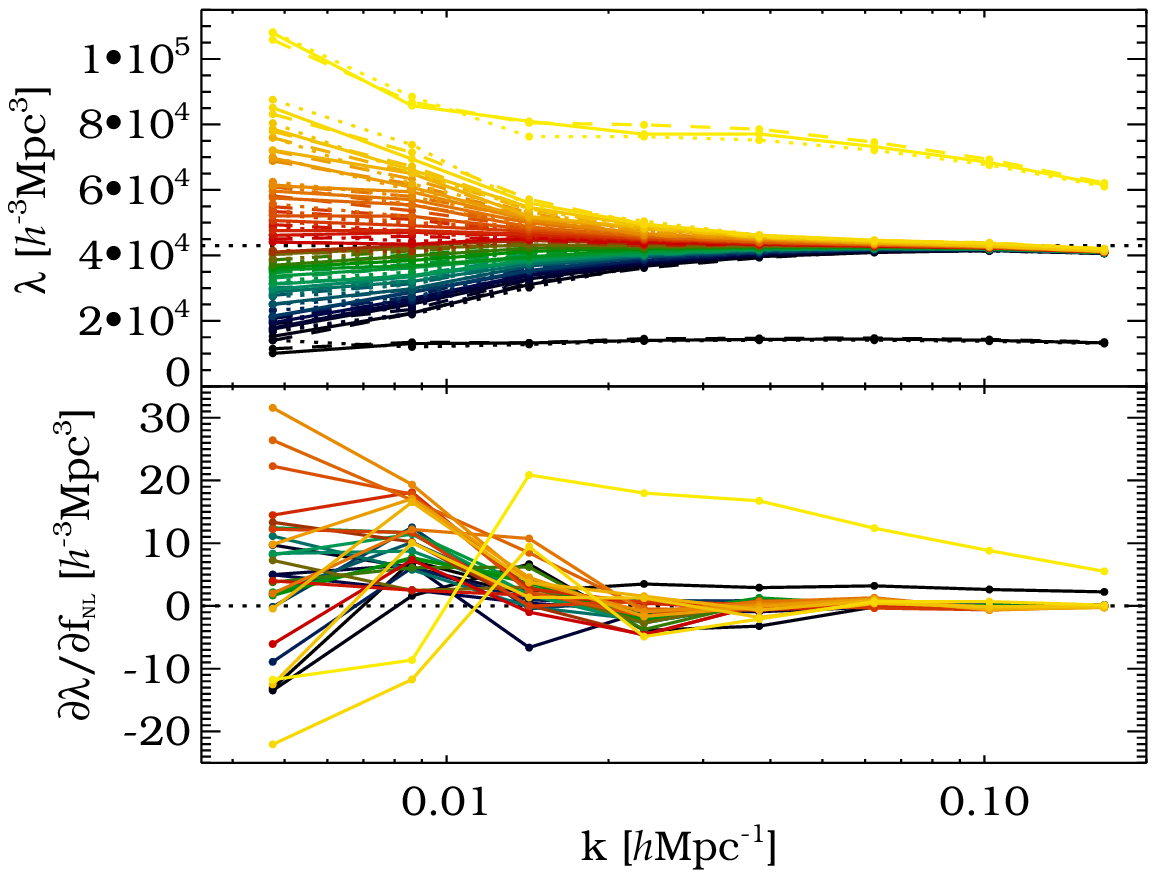}
\includegraphics{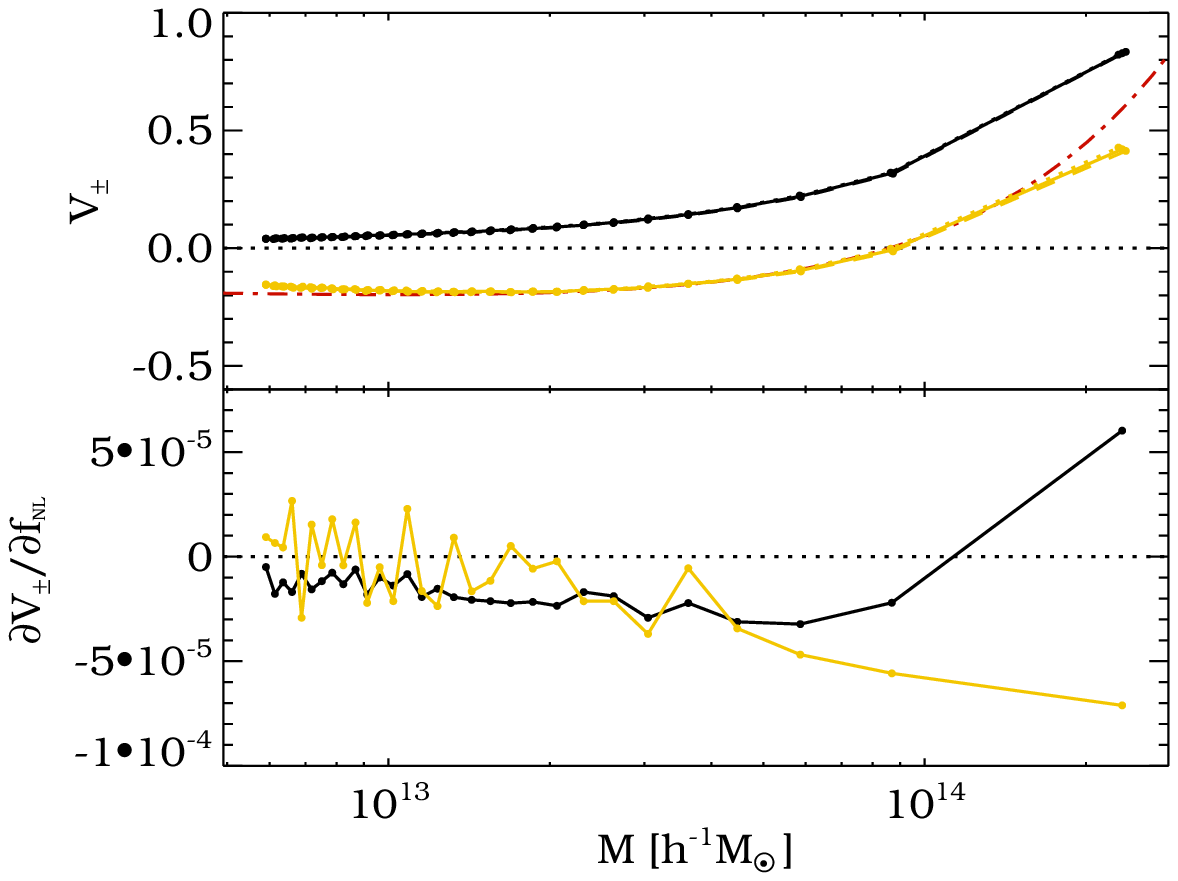}}
\caption{Eigenvalues (left panel) and eigenvectors (right panel) of the shot noise matrix $\E$ for $\fnl=0$ (solid), $+100$ (dashed) and $-100$ (dotted) in the case of $30$ mass bins. Their derivatives with respect to $\fnl$ are plotted underneath. For clarity, only the two eigenvectors $\Vpm$ along with their derivatives are shown in the right panel. The straight dotted line in the upper left panel depicts the value $1/\bar{n}$ and the red (dot-dashed) curve in the top right panel shows $b_2(M)$ computed from the peak-background split formalism, scaled to the value of $\Vp$ at $M\simeq3\times10^{13}\hMsun$. Results are shown for FOF halos at $z=0$.}
\label{SN}
\end{figure*}

The shot noise matrix $\E$ has been studied using simulations with Gaussian initial conditions in \cite{Hamaus2010}. Figure~\ref{SN} displays the eigenstructure of this matrix for $\fnl=0$ (solid curves) and $\fnl=\pm 100$ (dashed and dotted curves). The left panel depicts all the eigenvalues (top) and their derivatives with respect to $\fnl$ (bottom), while the right panel shows the two important eigenvectors $\Vp$ and $\Vm$ (top) along with their derivatives (bottom). The eigenstructure of $\E$ is accurately described by Eq.~(\ref{E_eb}), even in the non-Gaussian case. Namely, we still find one enhanced eigenvalue $\lambda_+$ and one suppressed eigenvalue $\lambda_-$. The remaining $N-2$ eigenvalues $\lambda_{\mathrm{P}}^{\left(N-2\right)}$ are degenerate with the value $1/\bar{n}$, the Poisson noise expectation. This means that our Gaussian bias-expansion model from Eq.~(\ref{model}) still works to describe $\E$ in the weakly non-Gaussian regime.

Note however that, owing to sampling variance, the decomposition into eigenmodes becomes increasingly noisy towards larger scales. This leads to an artificial breaking of the eigenvalue degeneracy which manifests itself as a scatter around the mean value $1/\bar{n}$. This scatter is the major contribution of sampling variance in the halo covariance matrix $\C$. Although we can eliminate most of it by setting $\lambda_{\mathrm{P}}^{\left(N-2\right)}\equiv1/\bar{n}$, a residual degree of sampling variance will remain in $\lambda_+$ and $\lambda_-$, as well as in $\bg$ and $P$.

As is apparent from the left panel in Fig.~\ref{SN}, the dominant eigenvalue $\lambda_+$ exhibits a small, but noticeable $\fnl$-dependence similar to that of $\mathcal{E}_{\delta^2}$ in Fig.~\ref{sn_m2}, which is about $2\%$ in this case. Its derivative, $\partial\lambda_+/\partial\fnl$, clearly dominates the derivative of all other eigenvalues (which are all consistent with zero due to matched abundances). Only the derivative of the suppressed eigenvalue $\lambda_-$ shows a similar $\fnl$-dependence of $\sim2\%$, albeit at a much lower absolute amplitude. To check the convergence of our results, we repeated the analysis with $100$ and $200$ bins and found both derivatives of $\lambda_+$ and $\lambda_-$ to increase, supporting an $\fnl$-dependence of these eigenvalues.

By contrast, the eigenvectors $\Vp$ and $\Vm$ shown in the right panel of Fig.~\ref{SN} exhibit very little dependence on $\fnl$ (the different lines are all on top of each other). The derivatives of $\Vp$ and $\Vm$ with respect to $\fnl$ shown in the lower panel reveal a very weak sensitivity to $\fnl$ which is less than $0.5\%$ for most of the mass bins (for the most massive bin it reaches up to $1\%$). We repeated the same analysis with $100$ and $200$ mass bins and found that the relative differences between the measurements in Gaussian and non-Gaussian simulations further decrease. We thus conclude that the eigenvectors $\Vp$ and $\Vm$ can be assumed independent of $\fnl$ to a very high accuracy.

Our findings demonstrate that the two-point statistics of halos are sensitive to primordial non-Gaussianity beyond the linear-order effect of Eq.~(\ref{b(k,fnl)}) derived in \cite{Dalal2008,Matarrese2008,Slosar2008}. However, the corrections are tiny if one considers a single bin containing many halos of very different mass (see \cite{Kendrick2010}) due to mutual cancellations from $b_2$-terms of opposite sign. Only two specific eigenmodes of the shot noise matrix (corresponding to two different weightings of the halo density field) inherit a significant dependence on $\fnl$. This is most prominently the case for the eigenmode corresponding to the highest eigenvalue $\lambda_+$. Its eigenvector, $\Vp$, is shown to be closely related to the second-order bias $\bg_2$ in Eq.~(\ref{E-model}). As can be seen in the upper right panel of Fig.~\ref{SN}, $\Vp$ measured from the simulations, and the function $b_2(M)$ calculated from the peak-background split formalism \cite{Sheth1999,Scoccimarro2001}, agree closely (note that $b_2(M)$ has been rescaled to match the normalized vector $\Vp$).

In the continuous limit this implies that weighting the halo density field with $b_2(M)$ selects the eigenmode with eigenvalue $\lambda_+$ given in Eq.~(\ref{lambda-model}). Since $\lambda_+$ depends on $\fnl$ through the quantity $\mathcal{E}_{\dm^2}$ defined in Eq.~(\ref{sigma_dm2}), the resulting weighted field will show the same $\fnl$-dependence. However, this $\fnl$-dependence cannot immediately be exploited to constrain primordial non-Gaussianity, because the Fourier modes of $\mathcal{E}_{\dm^2}$ are heavily correlated due to the convolution of $\delta$ with itself in Eq.~(\ref{sigma_dm2}), and thus do not contribute to the Fisher information independently. The bottom line is that for increasingly massive halo bins with large $b_2$, the term $\mathcal{E}_{\dm^2}$ makes an important contribution to the halo power spectrum \emph{and} shows a significant dependence on $\fnl$. It is important to take into account this dependence when attempting to extract the best-fit value of $\fnl$ from high-mass clusters, so as to avoid a possible measurement bias.
Although it provides some additional information on $\fnl$, we will ignore it in the following and quote only lower limits on the Fisher information content.

\subsection{Constraints from Halos and Dark Matter \label{sec:halos&dm}}
Let us first assume the underlying dark matter density field $\dm$ is available in addition to the galaxy distribution. Although this can in principle be achieved with weak-lensing surveys using tomography, the spatial resolution will not be comparable to that of galaxy surveys. To mimic the observed galaxy distribution we will assume that each dark matter halo (identified in the numerical simulations) hosts exactly one galaxy. A further refinement in the description of galaxies can be accomplished with the specification of a halo occupation distribution for galaxies \cite{Berlind2002,Cai2011}, but we will not pursue this here. Instead, we can think of the halo catalogs as a sample of central halo galaxies from which satellites have been removed. We also neglect the effects of baryons on the evolution of structure formation, which are shown to be marginally influenced by primordial non-Gaussianity at late times \cite{Maio2011}.

\subsubsection{Single tracer: uniform weighting}
In the simplest scenario we only consider one single halo mass bin. In this case, all the observed halos (galaxies) of a survey are correlated with the underlying dark matter density field in Fourier space to determine their scale-dependent effective bias, which can then be compared to theoretical predictions. In practice, this translates into fitting our theoretical model for the scale-dependent effective bias, Eq.~(\ref{b(k,fnl)}), to the Fourier modes of the density fields and extracting the best fitting value of $\fnl$ together with its uncertainty. For a single halo mass bin, we can employ Eq.~(\ref{chi2m_1}) and sum over all the Fourier modes.

In the Gaussian simulations, we measure the scale-independent effective bias $b_{\mathrm{G}}$ via the estimator $\langle\dhh\dm\rangle/\langle\dm^2\rangle$ and the shot noise $\mathcal{E}$ via $\langle\left(\dhh-b_{\mathrm{G}}\dm\right)^2\rangle$, and average over all modes with $k\le0.032h\mathrm{Mpc}^{-1}$. In practice, $b_{\mathrm{G}}$ and $\mathcal{E}$ are not directly observable, but a theoretical prediction based on the peak-background split \cite{Sheth1999,Scoccimarro2001} and the halo model \cite{Hamaus2010} provides a reasonable approximation to the measured $b_{\mathrm{G}}$ and $\mathcal{E}$, respectively, (see Sec.~\ref{sec:HM}). Note that for bins covering a wide range of halo masses, the $\fnl$-dependence of the shot noise is negligible \cite{Kendrick2010} and it is well approximated by its Gaussian expectation.

\begin{figure*}[!t]
\centering
\resizebox{\hsize}{!}{
\includegraphics{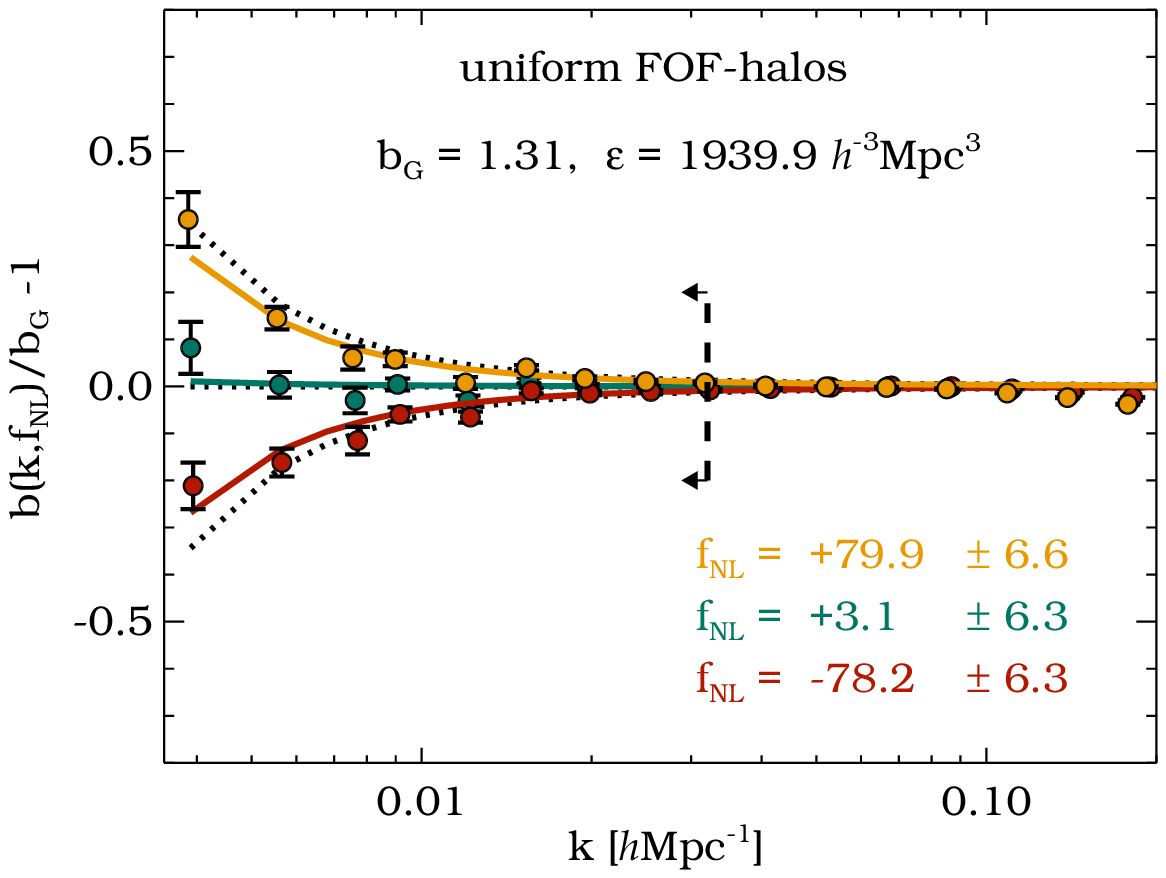}
\includegraphics{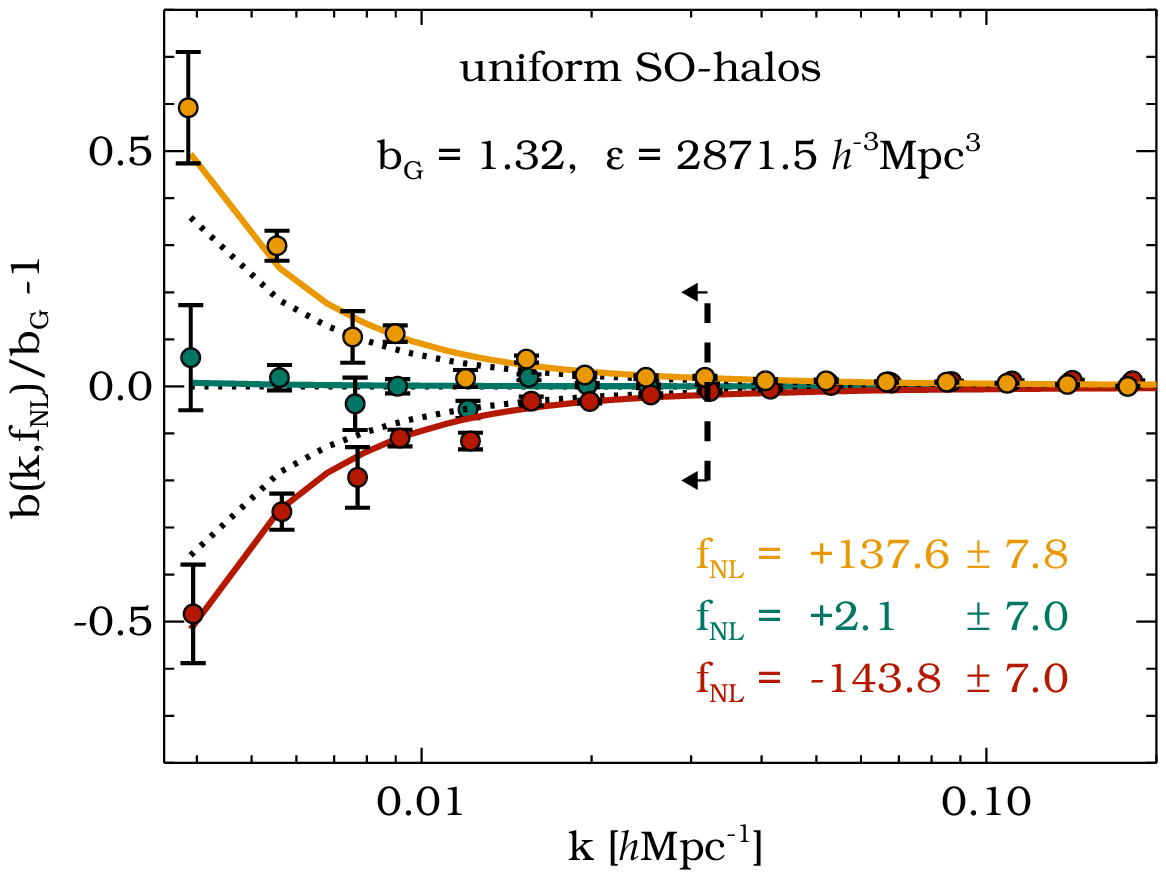}}
\caption{Relative scale dependence of the effective bias from all FOF (left panel) and SO halos (right panel) resolved in our N-body simulations ($M_{\mathrm{min}}\simeq 5.9\times10^{12}\hMsun$), which are seeded with non-Gaussian initial conditions of the local type with $\fnl=+100,0,-100$ (solid lines and data points from top to bottom). The solid lines show the best fit to the linear theory model of Eq.~(\ref{b(k,fnl)}), taking into account all the modes to the left of the arrow. The corresponding best-fit values are quoted in the bottom right of each panel. The dotted lines show the model evaluated at the input values $\fnl=+100,0,-100$. The results assume knowledge of the dark matter density field and an effective volume of $V_{\mathrm{eff}}\simeq50h^{-3}\mathrm{Gpc}^3$ at $z=0$.}
\label{fit_u}
\end{figure*}

Figure~\ref{fit_u} shows the best fits of Eq.~(\ref{b(k,fnl)}) to the simulations with $\fnl=0,\pm100$ using all the halos of our FOF (left panel) and SO catalogs (right panel). In order to highlight the relative influence of $\fnl$ on the effective bias, we normalize the measurements by the large-scale Gaussian average $b_{\mathrm{G}}$ and subtract unity. The resulting best-fit values of $\fnl$ along with their one-sigma errors are quoted in the lower right for each case of initial conditions. The $68\%$-confidence region is determined by the condition $\Delta\chi^2(\fnl)=1$. Note that we include only Fourier modes up to $k\simeq 0.032h\mathrm{Mpc}^{-1}$ in the fit, as linear theory begins to break down at higher wave numbers.

\begin{figure*}[!t]
\centering
\resizebox{\hsize}{!}{
\includegraphics{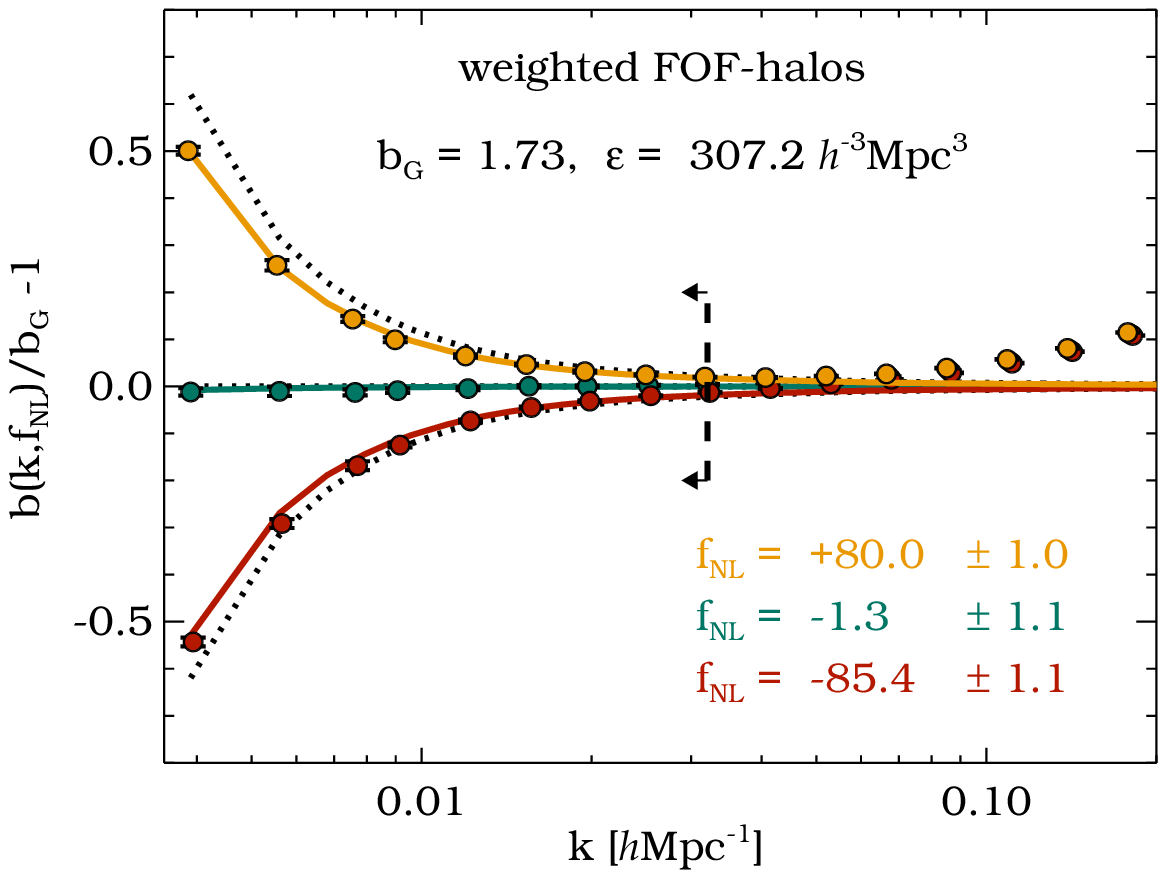}
\includegraphics{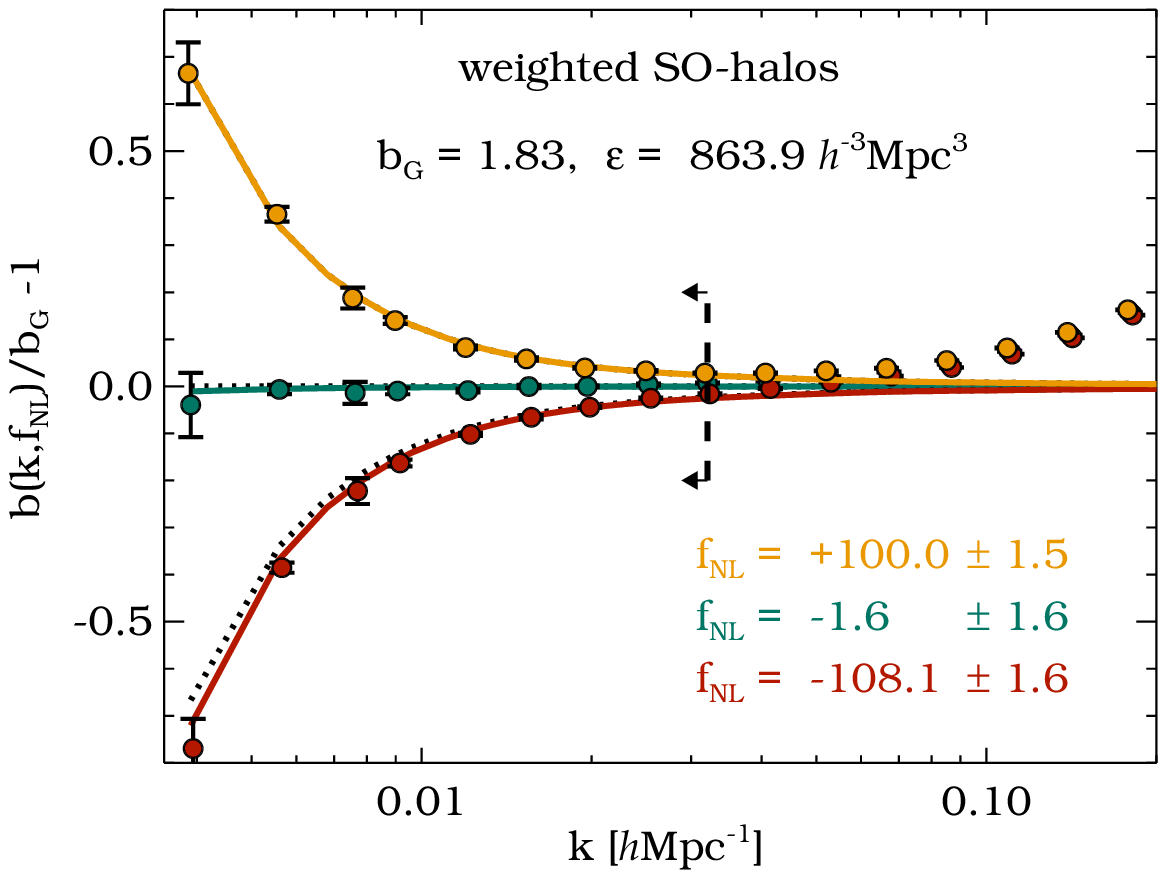}}
\caption{Same as Fig.~\ref{fit_u}, but for weighted halos that have minimum stochasticity relative to the dark matter. Note that the one-sigma errors on $\fnl$ are reduced by a factor of $\sim 5$ compared to uniform weighting. In the case of SO halos the input values for $\fnl$ are well recovered by the best-fit, while FOF halos still show a suppression of $\sim20\%$ ($q\simeq0.8$) in the best-fit $\fnl$.}
\label{fit_w}
\end{figure*}

Obviously, the best-fit values for $\fnl$ measured from the FOF halo catalogs are about $20\%$ below the input values. A suppression of the non-Gaussian correction to the bias of FOF halos has already been reported by \cite{Grossi2009,Pillepich2010}. These authors showed that the replacement $\dc\rightarrow q\dc$ with $q=0.75$ in Eq.~(\ref{b(k,fnl)}) yields a good agreement with their simulation data. In our framework, including this ``$q$-factor'' is equivalent to exchanging $\fnl\rightarrow\fnl/q$ and $\sigma_{\fnl}\rightarrow\sigma_{\fnl}/q$, owing to the linear scaling of Eq.~(\ref{b(k,fnl)}) with $\dc$. Repeating the chi-square minimization with $q=0.75$ yields best-fit values that are consistent with our input values, namely $\fnl=+107.0\pm8.3$, $+1.8\pm8.7$ and $-104.0\pm8.5$. In fact, the closest match to the input $\fnl$-values is obtained for a slightly larger $q$ of $\simeq0.8$.

Note that \cite{Grossi2009} attributed this suppression to ellipsoidal collapse. However, this conclusion seems rather unlikely since ellipsoidal collapse increases the collapse threshold or, equivalently, implies $q>1$ \cite{Sheth2001}. A more sensible explanation arises from the fact that a linking length of 0.2 times the mean interparticle distance can select regions with an overdensity as low as $\Delta\sim1/0.2^3=125$ (with respect to the mean background density $\rhom$), which is much less than the virial overdensity $\Dc(z=0)\simeq340$ associated with a linear overdensity $\dc$ (see \cite{Eke1996,Valageas2010,More2011}). Therefore, we may reasonably expect that, on average, FOF halos with this linking length trace linear overdensities of height less than $\dc$.

In the case of SO halos, however, we observe the opposite trend. As is apparent in the right panel of Fig.~\ref{fit_u}, the model from Eq.~(\ref{b(k,fnl)}) overestimates the amplitude of primordial non-Gaussianity by roughly $40\%$. This is somewhat surprising since the overdensity threshold $\Delta_c\simeq 340$ used to identify the SO halos at $z=0$ is precisely the virial overdensity predicted by the spherical collapse of a linear perturbation of height $\dc$. As we will see shortly, however, an optimal weighting of halos can remove this overshoot and therefore noticeably improve the agreement between model and simulations.

\subsubsection{Single tracer: optimal weighting}

As demonstrated in \cite{Hamaus2010}, the shot noise matrix $\E$ exhibits nonzero off-diagonal elements from correlations
between halos of different mass. Thus, in order to extract the full information on halo statistics, it is necessary to include these correlations into our analysis. For this purpose, we must employ the more general chi-square of Eq.~(\ref{chi2m}). The halo density field is split up into $N$ consecutive mass bins in order to construct the vector $\dhalo$, and the full shot noise matrix $\E$ must be considered.

However, this approach can be simplified, since we know that $\E$ exhibits one particularly low eigenvalue~$\lambda_-$. Because the Fisher information content on $\fnl$ from Eq.~(\ref{F_m_text}) is proportional to the inverse of $\E$ (this is true at least for the dominant part $\gamma$), it is governed by the eigenmode corresponding to this eigenvalue. In \cite{Hamaus2010} it has been shown that this eigenmode dominates the clustering signal-to-noise ratio. In the continuous limit (infinitely many bins), it can be projected out by performing an appropriate weighting of the halo density field. The corresponding weighting function, denoted as \emph{modified mass weighting} with functional form
\begin{equation}
w(M)=M+M_0 \;, \label{w(M)}
\end{equation}
was found to minimize the stochasticity of halos with respect to the dark matter. Here, $M$ is the individual halo mass and $M_0$ a constant whose value depends on the resolution of the simulation. It is approximately $3$ times the minimum resolved halo mass $M_{\mathrm{min}}$, so in this case $M_0\simeq1.8\times10^{13}\hMsun$. The weighted halo density field is computed as
\begin{equation}
\delta_w=\frac{\sum_i w(M_i)\delta_{\mathrm{h}_i}}{\sum_i w(M_i)}\equiv\frac{\boldsymbol{w}^\intercal\dhalo}{\boldsymbol{w}^\intercal\openone}\;, \label{weight}
\end{equation}
where we have combined the weights of the individual mass bins into a vector $\boldsymbol{w}$ in the last expression. Because the chi-square in Eq.~(\ref{chi2m}) is dominated by only one eigenmode, it simplifies to the form of Eq.~(\ref{chi2m_1}) with the halo field $\dhh$ being replaced by the weighted halo field~$\delta_w$. Note also that $b_{\mathrm{G}}$ and $\mathcal{E}$ have to be replaced by the corresponding weighted quantities (see \cite{Hamaus2010}).

The results are shown in Fig.~\ref{fit_w} for both FOF and SO halos. We observe a remarkable reduction in the error on $\fnl$ by a factor of $\sim 4-6$ (depending on the halo finder) when replacing the uniform sample used in Fig.~\ref{fit_u} by the optimally weighted one. While for the FOF halos the predicted amplitude of the non-Gaussian correction to the halo bias still shows the $20\%$ suppression (again, this can be taken into account by introducing a $q$-factor into our fit), for the SO halos the best-fit values of $\fnl$ now agree much better with the input values, i.e., $q\simeq1$.

Therefore, the large discrepancy seen in Fig.~\ref{fit_u} presumably arises from noise in the SO mass assignment at low mass. To ascertain whether this is the case, we repeat the analysis, increasing the threshold for the minimum number of particles per halo and discarding all halos below that threshold. If this threshold reaches $40$ particles per halo, we find the best-fit $\fnl$ to be much closer to the input values, namely $\fnl=+102.6\pm4.6$, $+2.1\pm4.5$ and $-103.5\pm4.4$. This suggests that most of the discrepancy seen in the right panel of Fig.~\ref{fit_u} is due to poorly resolved halos of mass $M\lesssim2M_\mathrm{min}$ \cite{Knebe2011}. However, modified mass weighting removes this discrepancy since halos at low mass are given less weight.

\begin{figure*}[!t]
\centering
\resizebox{\hsize}{!}{
\includegraphics{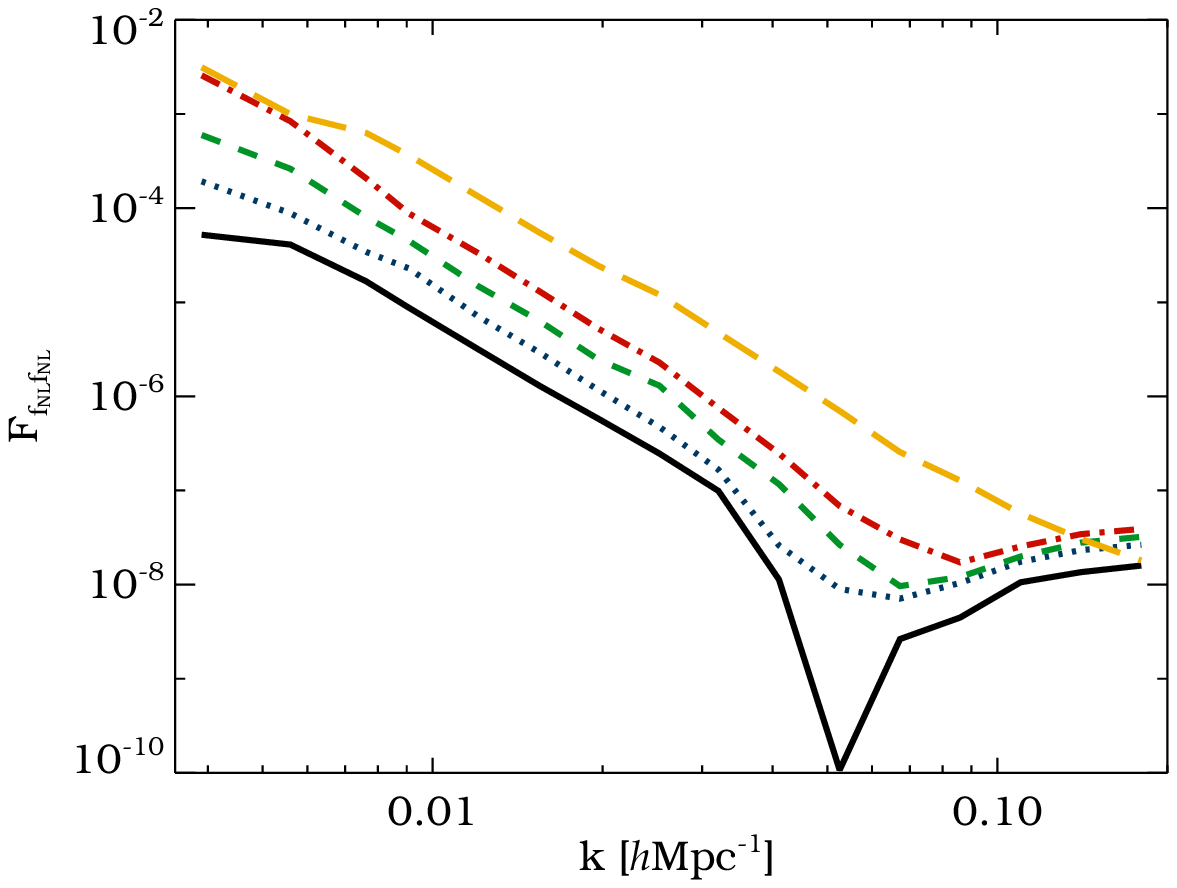}
\includegraphics{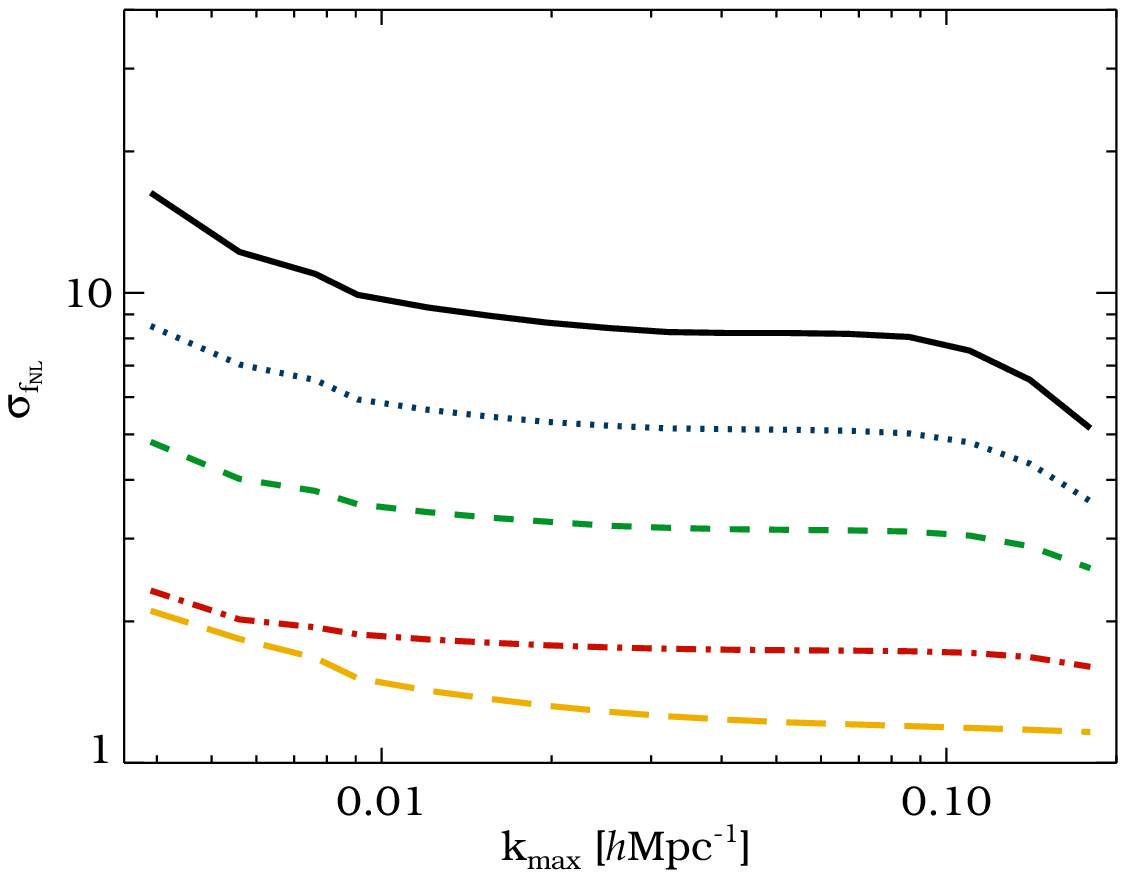}}
\caption{Fisher information (left panel) and one-sigma error on $\fnl$ (right panel, $k_{\mathrm{min}}=0.0039h\mathrm{Mpc}^{-1}$, $V_{\mathrm{eff}}\simeq50h^{-3}\mathrm{Gpc}^3$) from simulations of FOF halos and dark matter at $z=0$. The lines show the results for $1$ (solid black), $3$ (dotted blue), $10$ (dashed green) and $30$ (dot-dashed red) uniform halo mass bins, as well as for $1$ weighted bin (long-dashed yellow).}
\label{fisher_m}
\end{figure*}

Our findings are consistent with the ones of \cite{Desjacques2009}, where the non-Gaussian bias of SO halos has been measured also at higher redshifts and mass thresholds, and the results of \cite{Desjacques2010b}, where the fractional deviation from the Gaussian mass function for both FOF and SO halos was presented (see their Fig.~5). The remarkable improvement in the constraints on $\fnl$ follows from the fact that the stochasticity (shot noise) of the optimally weighted halo density field is strongly suppressed with respect to the dark matter \cite{Hamaus2010}. This means that the fluctuations of the halo and the dark matter overdensity fields are more tightly correlated and the variance of the estimator $\langle\dhh\dm\rangle/\langle\dm^2\rangle$ for the effective bias is minimized. Also, cosmic variance fluctuations inherent in both $\dm$ and $\dhh$ are canceled in this ratio (see Appendix \ref{appendix3}). Since the scale dependence of this estimator is a direct probe of primordial non-Gaussianity, the error on $\fnl$ is significantly reduced. At the same time, modified mass weighting increases the magnitude of $b_{\mathrm{G}}$. We will show below that the constraints on $\fnl$ are indeed optimized with this approach.

Finally, we can test our assumption about the likelihood function as defined in Eq.~(\ref{likelihood}) being of a Gaussian form and thus yielding the correct Fisher information. Non-Gaussian corrections could arise from correlated $k$-modes in the covariance matrix (as present in the eigenmode $\lambda_+$ of the shot noise matrix), preventing the Fisher information from being a single integral over $k$. The error $\sigma_b$ on the effective bias in Figs.~\ref{fit_u} and \ref{fit_w} is determined from the variance amongst our sample of $12$ realizations and thus provides an independent way of testing the value for $\sigma_{\fnl}$: from Eq.~(\ref{b(k,fnl)}) we can determine $\sigma_{\fnl}=\sigma_b/(b_{\mathrm{G}}-1)u(k,z)$ and compare it to the value obtained from the chi-square fit with Eq.~(\ref{chi2m}). Applying the two methods, we find no significant differences in $\sigma_{\fnl}$, so at least up to the second moment of the likelihood function, the assumption of it being Gaussian seems reasonable for the considered values of $\fnl$.

\subsubsection{Multiple tracers}
Let us now estimate the minimal error on $\fnl$ achievable with a given galaxy survey for the general case, dividing halos into multiple mass bins. The Fisher information is given by Eq.~(\ref{F_m_text}) or~(\ref{F_lin}), depending on whether the dark matter density field is known or not, and the minimal error on $\fnl$ is determined via integration over all observed modes in the volume $V$,
\begin{equation}
\sigma_{\fnl}^{-2}=\frac{V}{2\pi^2}\int_{k_{\mathrm{min}}}^{k_{\mathrm{max}}}F_{\fnl\fnl}(k)\;k^2\mathrm{d}k \; . \label{sigma_fnl}
\end{equation}
The largest modes with wave number $k_{\mathrm{min}}=2\pi/L_{\mathrm{box}}\simeq0.0039h\mathrm{Mpc}^{-1}$ available from our $N$-body simulations are smaller than the largest modes in a survey of $50h^{-3}\mathrm{Gpc}^3$ volume ($k_{\mathrm{min}}\simeq0.0017h\mathrm{Mpc}^{-1}$ ), since we only obtain an \emph{effective} volume by considering $12$ smaller simulation boxes. Because the signal from $\fnl$ is strongest at low $k$, our results slightly underestimate the total Fisher information. However, we can roughly estimate that on larger scales ($k_{\mathrm{min}}<0.0039h\mathrm{Mpc}^{-1}$), $F_{\fnl\fnl}(k)\sim u^2(k)P(k)\sim k^{-4}k^{n_s}$ [see Eqs.~(\ref{u(k,z)}), (\ref{F_lin}) and (\ref{F_m_text}), as well as Figs.~\ref{fisher_m} and \ref{fisher_h}], and thus $\sigma_{\fnl}\sim  \ln\left(k_{\mathrm{max}}/k_{\mathrm{min}}\right)^{-1/2}$ assuming $n_s\simeq1$, a relatively weak dependence on $k_{\mathrm{min}}$. In our case this amounts to an overestimation of $\sigma_{\fnl}$ by roughly $20\%$.

Note that we only consider the $\fnl$-$\fnl$-element of the Fisher matrix. In principle we would have to consider various other parameters of our cosmology and then marginalize over them, i.e., compute $\left(F^{-1}\right)_{\fnl\fnl}$ \cite{Carbone2010}. However, any degeneracy with cosmological parameters is largely eliminated when multiple tracers are considered, since the underlying dark matter density field mostly cancels out in this approach \cite{Seljak2009a}. A mathematical demonstration of this fact is presented in Appendix \ref{appendix3}.

Recent studies have developed a gauge-invariant description of the observable large-scale power spectrum consistent with general relativity \cite{Yoo2009a,Yoo2009b,Yoo2010,McDonald2009b,Bonvin2011,Challinor2011,Baldauf2011b}. In particular, it has been noted that the general relativistic corrections to the usually adopted Newtonian treatment leave a signature in the galaxy power spectrum that is very similar to the one caused by primordial non-Gaussianity of the local type \cite{Wands2009,Bruni2011,Jeong2011}. However, in a multitracer analysis the two effects can be distinguished sufficiently well, so that the ability to detect primordial non-Gaussianity is little compromised in the presence of general relativistic corrections \cite{Yoo2011}.

In order to make the most conservative estimates we will discard all the terms featuring $\E'$ in the Fisher matrix, since it is not obvious how much information on $\fnl$ can actually be extracted from the shot noise matrix. $\E$~is indeed close to a pure white-noise quantity and we find its Fourier modes to be highly correlated. Therefore, in order to extract residual information on $\fnl$, one would have to decorrelate those modes through an inversion of the correlation matrix among $k$-bins (see \cite{Kiessling2011}). However, in light of the limited volume of our simulations this can be a fairly noisy procedure, especially when the halo distribution is additionally split into narrow mass bins. Hence, for the Fisher information content on $\fnl$ assuming knowledge of both halos and dark matter, we will retain only the first term in Eq.~(\ref{F_m_text}) and provide a lower limit:
\begin{equation}
F_{\fnl\fnl}\ge\gamma\equiv\bg'^\intercal\E^{-1}\bg' P \; . \label{F_fnl}
\end{equation}
To calculate $F_{\fnl\fnl}$, we measure the functions $\bg(k)$, $\bg'(k)$, $\E(k)$ and $P(k)$ from our $N$-body simulations (see Figs.~\ref{sn_m2}, \ref{bias} and \ref{SN}). In order to mitigate sampling variance in the multibin case, we then use Eq.~(\ref{E_eb}) to recalculate the shot noise matrix. Namely, we set all the eigenvalues $\lambda_{\mathrm{P}}^{\left(N-2\right)}$ equal to the average value $1/\bar{n}$, and measure $\lambda_+$, $\lambda_-$, as well as $\Vp$ and $\Vm$ directly from the numerical eigendecomposition of $\E$.

Figure~\ref{fisher_m} depicts $F_{\fnl\fnl}(k)$ and $\sigma_{\fnl}(k_\mathrm{max})$ with fixed $k_\mathrm{min}=0.0039h\mathrm{Mpc}^{-1}$ for the cases of $1$, $3$, $10$ and $30$ halo mass bins. Clearly, the finer the sampling into mass bins, the higher the information content on $\fnl$. The weighted halo density field with minimal stochasticity relative to the dark matter (corresponding to a continuous sampling of infinitely many bins) yields more than a factor of 6 reduction in $\sigma_{\fnl}$ when compared to a single mass bin of uniformly weighted halos. This improvement agrees reasonably well with that seen in Figs. \ref{fit_u} and \ref{fit_w}, although the estimates for $\sigma_{\fnl}$ are slightly larger than those we obtained from the fitting procedure. This may be expected, since we only obtain an upper limit on $\sigma_{\fnl}$ from Eqs.~(\ref{sigma_fnl}) and (\ref{F_fnl}).

The inflection around $k\sim0.1h\mathrm{Mpc}^{-1}$ in $F_{\fnl\fnl}$ and $\sigma_{\fnl}$ marks a breakdown of the linear model from Eq.~(\ref{b_fnl}). We should not trust our results too much at high wave number, where higher-order contributions to the non-Gaussian effective bias may become important. It should also be noted that the inflection disappears for the weighted field, suggesting numerical issues to be less problematic in that case.

Further improvements can be achieved when going to lower halo masses (see Sec.~\ref{sec:HM}): the error on $\fnl$ is proportional to the shot noise of the halo density field (Eq.~(\ref{F_m1})), which itself is a function of the minimum halo mass $M_{\mathrm{min}}$. References~\cite{Hamaus2010,Cai2011} numerically investigated the extent to which the shot noise depends on $M_{\mathrm{min}}$ and proposed a method based on the halo model for extrapolating it to lower mass. It predicts the shot noise of the weighted halo density field to decrease linearly with $M_{\mathrm{min}}$, anticipating about 2 orders of magnitude further reduction in $\mathcal{E}$ when resolving halos down to $M_{\mathrm{min}}\simeq10^{10}\hMsun$. In terms of $\fnl$-constraints this is however somewhat mitigated by the fact that the Gaussian bias also decreases with $M_{\mathrm{min}}$, so the non-Gaussian correction to the effective bias in Eq.~(\ref{b(k,fnl)}) gets smaller. Furthermore, \cite{Hamaus2010} studied the effect of adding random noise to the halo mass (to mimic scatter between halo mass and the observables such as galaxy luminosity), while \cite{Cai2011} explored the redshift dependence of the optimally weighted halo density field and extended the method to halo occupation distributions for galaxies.

\begin{figure*}[!t]
\centering
\resizebox{\hsize}{!}{
\includegraphics{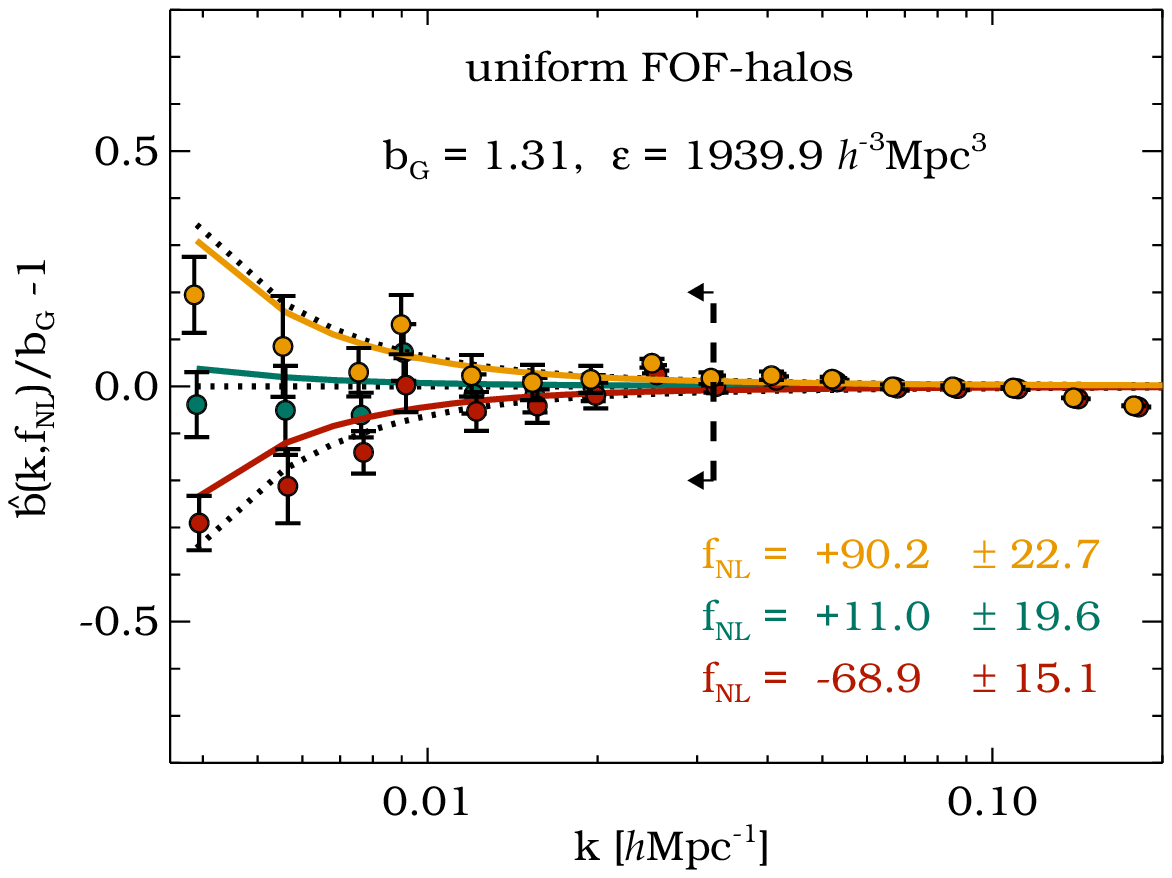}
\includegraphics{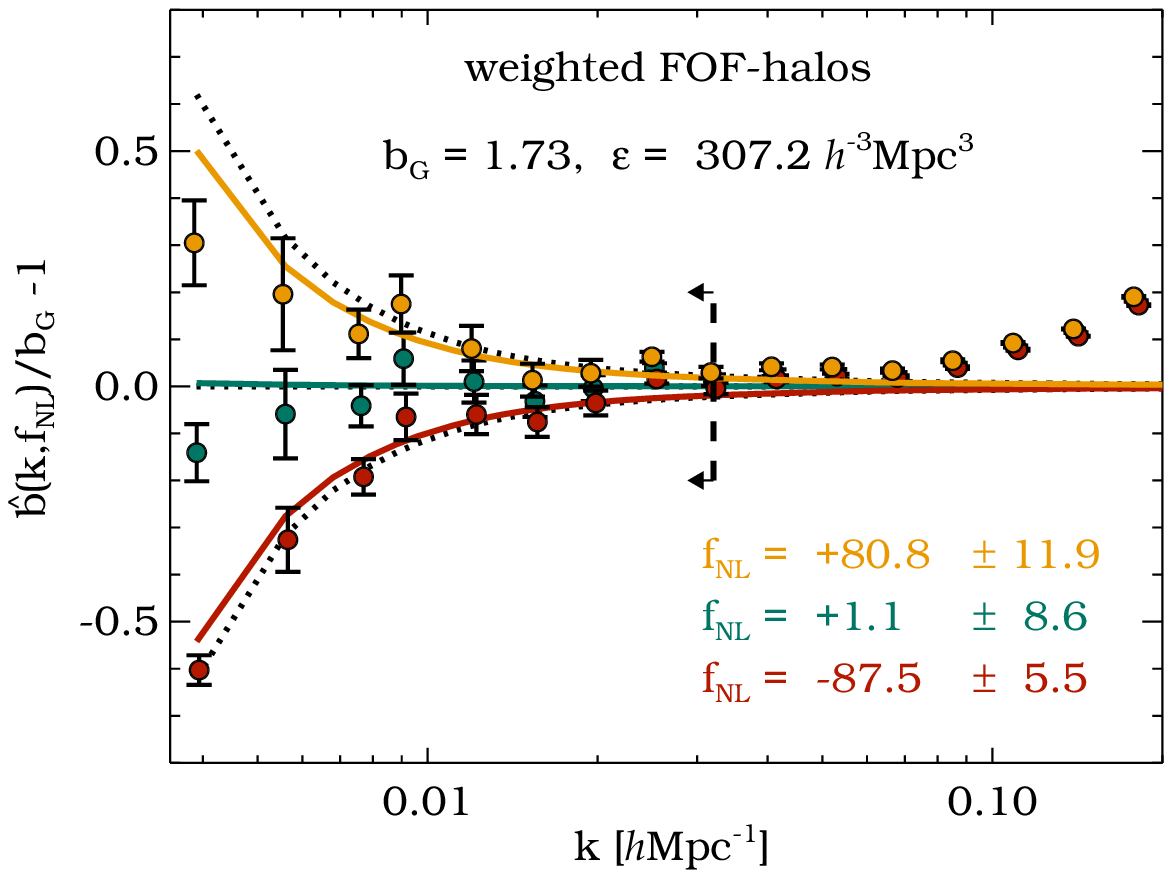}}
\caption{Relative scale dependence of the effective bias $\hat{b}$ estimated from all uniform (left panel) and weighted FOF halos (right panel) resolved in our N-body simulations ($M_{\mathrm{min}}\simeq 5.9\times10^{12}\hMsun$), which are seeded with non-Gaussian initial conditions of the local type with $\fnl=+100,0,-100$ (solid lines and data points from top to bottom). The solid lines show the best fit to the linear theory model of Eq.~(\ref{b(k,fnl)}), taking into account all the modes to the left of the arrow. The corresponding best-fit values are quoted in the bottom right of each panel. The dotted lines show the model evaluated at the input values $\fnl=+100,0,-100$. The results assume no knowledge of the dark matter density field and an effective volume of $V_{\mathrm{eff}}\simeq50h^{-3}\mathrm{Gpc}^3$ at $z=0$.}
\label{fit_hh}
\end{figure*}

\subsection{Constraints from Halos}
The scenario described above is optimistic in the sense that it assumes the dark matter density field is available. In the following section we will show that it is possible to considerably improve the constraints on $\fnl$ even without this
assumption. This is perhaps not surprising in light of the results in \cite{Hamaus2010,Cai2011}, where it was argued that halos can be used to reconstruct the dark matter to arbitrary precision, as long as they are resolved down to the required low-mass threshold.

\subsubsection{Single tracer}
Considering a single halo mass bin, we must again sum over all the Fourier modes in Eq.~(\ref{chi2_1}) and minimize this chi-square with respect to $\fnl$. Although we pretend to have no knowledge of the dark matter distribution, we determine $b_{\mathrm{G}}$ and $\mathcal{E}$ from our simulations. In realistic applications, however, these quantities will have to be accurately modeled. In addition, we use the linear power spectrum $P_0(k)$ instead of the simulated nonlinear dark matter power spectrum $P(k)$ in Eq.~(\ref{chi2_1}).

Since, in this case, we cannot determine the scale-dependent effective bias directly from the estimator $\langle\dhh\dm\rangle/\langle\dm^2\rangle$, we define the new estimator
\begin{equation}
\hat{b}\equiv\sqrt{\frac{\langle\dhh^2\rangle-\mathcal{E}}{P_0}} \;, \label{b_hat}
\end{equation}
which solely depends on the two-point statistics of halos. In Fig.~\ref{fit_hh} we plot this estimator together with the best-fit solutions for the scale-dependent effective bias obtained from the chi-square fit of Eq.~(\ref{chi2_1}). The left panel depicts the results obtained for uniform FOF halos. Compared to the previous case with dark matter, we observe the constraints on $\fnl$ to be weaker by a factor of $\sim3$. The main reason for this difference is the fact that sampling variance inherent in $\dhh$ is not canceled out by subtracting~$\dm$, as is done in Eq.~(\ref{chi2m_1}). This can also be seen in the estimator $\hat{b}$, where a division of the smooth linear power spectrum $P_0$ does not cancel the cosmic variance inherent in $\langle\dhh^2\rangle$. Hence, $\hat{b}$ shows significantly stronger fluctuations than $b=\langle\dhh\dm\rangle/\langle\dm^2\rangle$, which demonstrates how well the basic idea of sampling variance cancellation works.

\begin{figure*}[!t]
\centering
\resizebox{\hsize}{!}{
\includegraphics{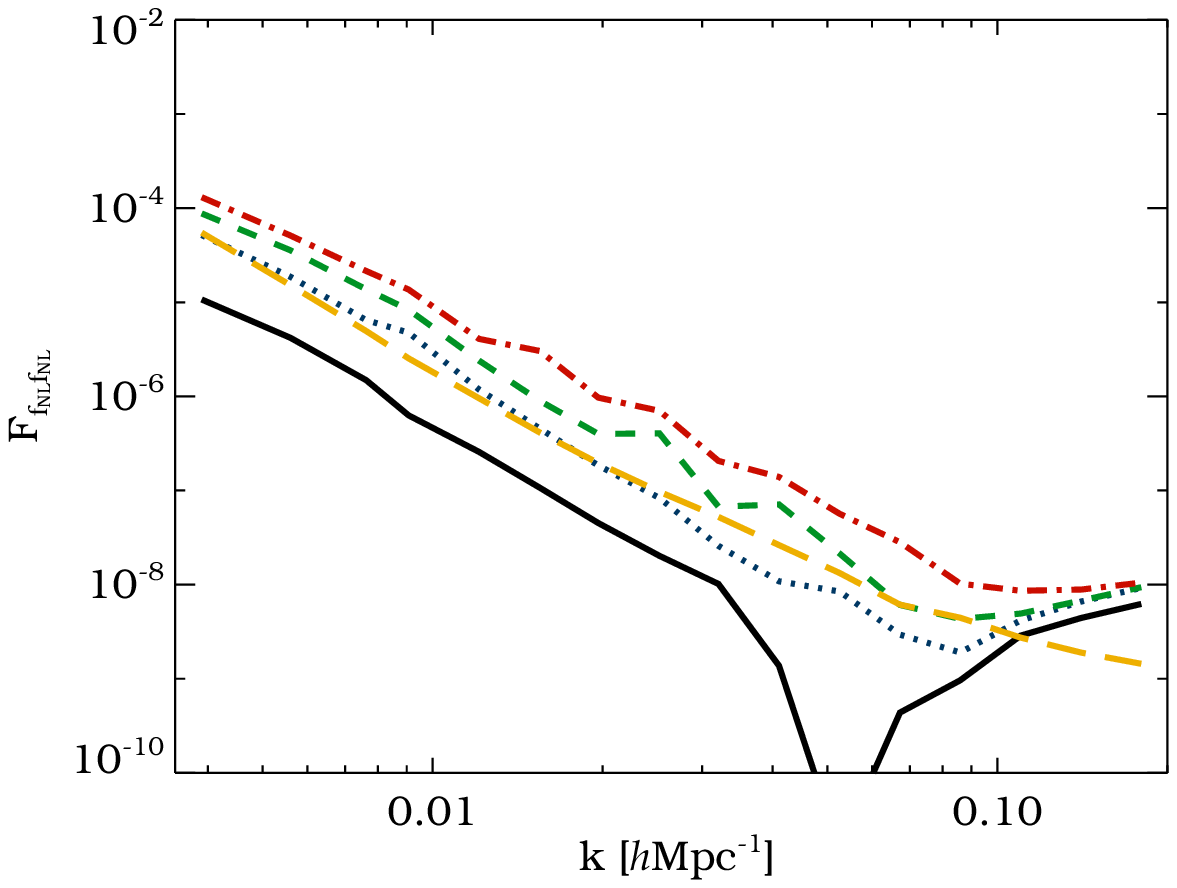}
\includegraphics{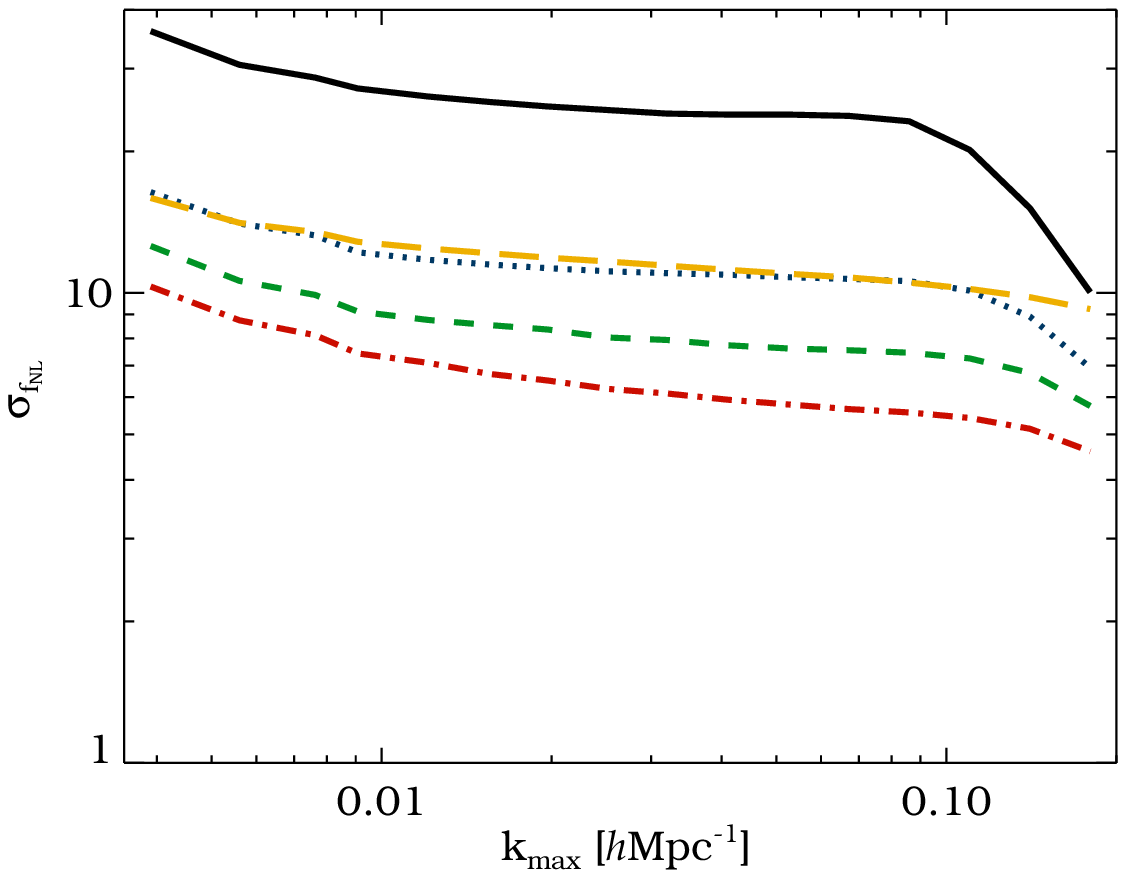}}
\caption{Fisher information (left panel) and one-sigma error on $\fnl$ (right panel, $k_{\mathrm{min}}=0.0039h\mathrm{Mpc}^{-1}$, $V_{\mathrm{eff}}\simeq50h^{-3}\mathrm{Gpc}^3$) from simulations of FOF halos only ($z=0$). The lines show the results for $1$ (solid black), $3$ (dotted blue), $10$ (dashed green) and $30$ (dot-dashed red) uniform halo mass bins, as well as for $1$ weighted bin (long-dashed yellow).}
\label{fisher_h}
\end{figure*}

Exchanging the uniform halo field $\dhh$ with the weighted one, $\delta_w$, the constraints on $\fnl$ improve by about a factor of $2-3$, as can be seen in the right panel of Fig.~\ref{fit_hh}. However, this improvement is mainly due to the larger value of $b_{\mathrm{G}}$ of the weighted sample, since the relative scatter among the data points remains unchanged. This is expected, because we do not consider a second tracer (e.g., the dark matter) in this case, and therefore do not cancel cosmic variance.

Comparing the uncertainty on $\fnl$ obtained from Eq.~(\ref{chi2}) with the one determined via the variance of $\hat{b}$ amongst our $12$ realizations, we can check once more the assumption of a Gaussian likelihood as given in Eq.~(\ref{likelihood}). Again, we find both methods to yield consistent values for $\sigma_{\fnl}$, suggesting any non-Gaussian corrections to the likelihood function to be negligible at this order.

\subsubsection{Multiple tracers}
If we want to exploit the gains from sampling variance cancellation in the case where the dark matter density field is not available, we have to perform a multitracer analysis of halos (see Appendix \ref{appendix3}), which is the focus of this section. We now consider Eq.~(\ref{chi2}) for the chi-square fit. In order to calculate the Fisher information, we use Eq.~(\ref{F_lin}) and thus neglect any possible contribution emerging from the $\fnl$-dependence of the shot noise matrix $\E$.

Numerical results for the Fisher information content and the one-sigma error on $\fnl$ are shown in Fig.~{\ref{fisher_h}} for $1$, $3$, $10$ and $30$ uniform FOF halo mass bins, as well as for $1$ weighted bin. Clearly, the cases of $10$ and $30$ uniform bins outperform a single bin of the weighted field in terms of Fisher information. This suggests that further improvements compared to the single weighted halo field can be achieved when all the correlations of sufficiently many halo mass bins are taken into account.

In principle, we want to split the halo density field into as many mass bins as possible and extrapolate the results to the limit of infinitely many bins (continuous limit). Note that in the high-sampling limit of $\bar{n}\rightarrow\infty$, $F_{\fnl\fnl}$ from Eq.~(\ref{F1_text}) is limited to $2\left(b'/b\right)^2$, whereas the same quantity for several mass bins, Eq.~(\ref{F_lin}), may surpass this bound (see Sec.~\ref{sec:HM}). In Sec.~\ref{sec:halos&dm} we showed that a single optimally weighted halo sample combined with the dark matter reaches the continuous limit in $F_{\fnl\fnl}$, which corresponds to a splitting into infinitely many bins in the multitracer approach. It is unclear, whether a similar goal can be achieved from halos alone, e.g., by considering two differently weighted tracers that would preserve all of the information on $\fnl$, because we do not know the continuous limit of the Fisher information in that case. We will therefore turn to theoretical predictions by the halo model in the following paragraph.

\section{Halo model predictions \label{sec:HM}}

\begin{figure*}[!t]
\centering
\resizebox{\hsize}{!}{
\includegraphics{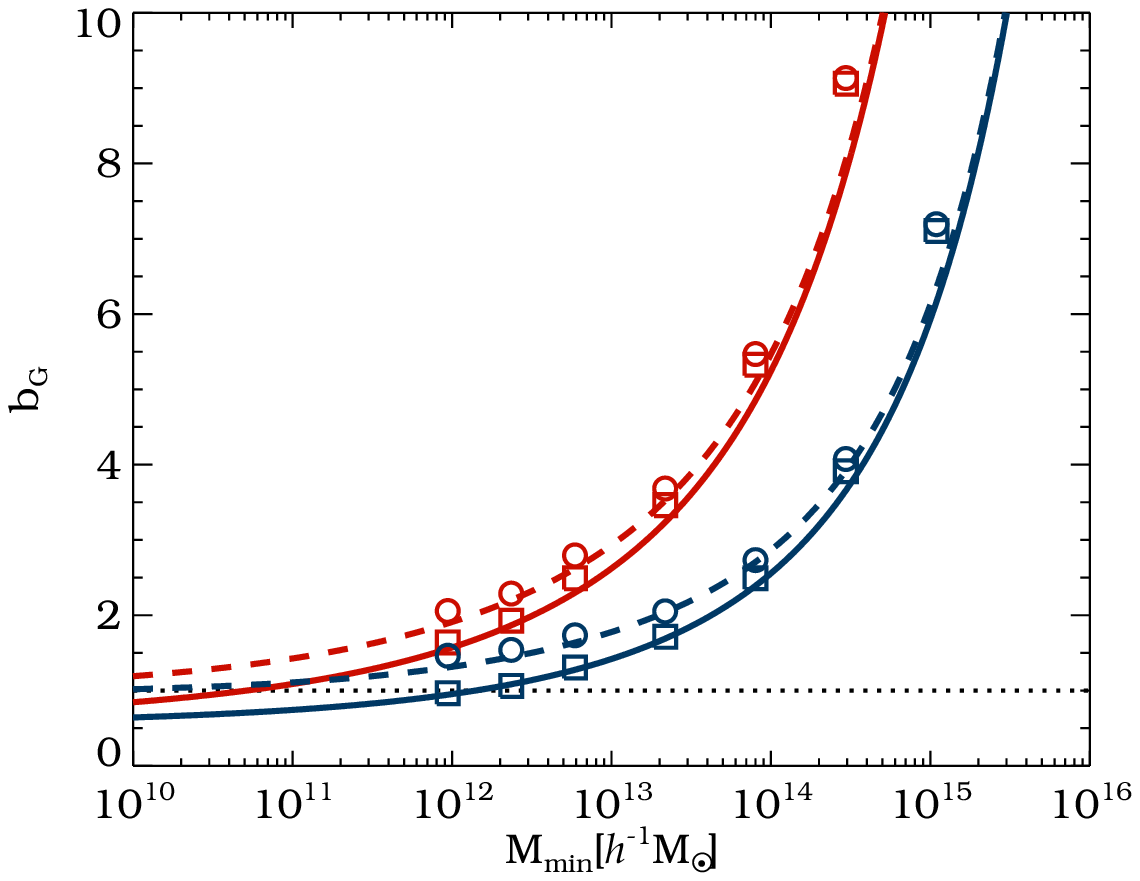}
\includegraphics{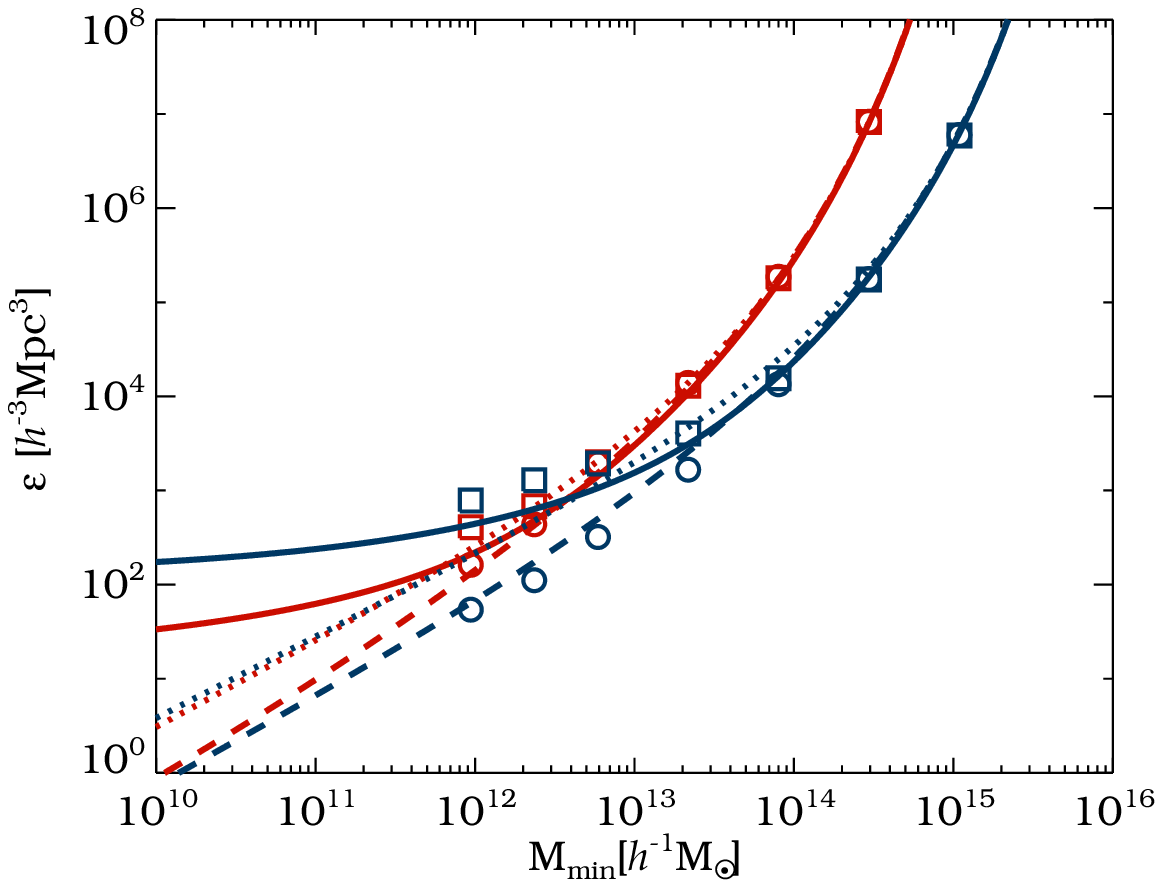}}
\caption{Halo model predictions for the mean scale-independent Gaussian bias (left panel) and shot noise (right panel) as a function of minimum halo mass from uniform- (solid lines) and weighted halos (dashed lines) from a single mass bin at $z=0$ (blue) and $z=1$ (red). $N$-body simulation results are overplotted, respectively, as squares and circles for different low-mass cuts. The dotted line in the left panel depicts $b_{\mathrm{G}}=1$, the ones in the right panel show the Poisson-model shot noise $\bar{n}_{\mathrm{tot}}^{-1}$.}
\label{HM_b_sn}
\end{figure*}

A useful theoretical framework for the description of dark matter and halo clustering is given by the halo model (see, e.g., \cite{Seljak2000}). Despite its limitations \cite{Crocce2008}, the halo model achieves remarkable agreement with the results from $N$-body simulations \cite{Seljak2000,Hamaus2010}. In particular, it provides an analytical expression for the shot noise matrix in the fiducial Gaussian case, given by
\begin{equation}
\E=\bar{n}^{-1}\I-\bg\Mr^\intercal-\Mr\bg^\intercal \;, \label{E_hm}
\end{equation}
where $\Mr\equiv\M/\rhom-\bg\langle nM^2\rangle/2\rhom^2$ and $\M$ is a vector containing the mean halo mass of each bin (see \cite{Hamaus2010} for the derivation). The Poisson model is recovered when we set $\Mr=0$. Here, $\bg$ can be determined by integrating the peak-background split bias $b(M)$ over the Sheth-Tormen halo mass function $dn/dM$ \cite{Sheth1999} in each mass bin. The expression $\langle nM^2\rangle/\rhom^2$ originates from the dark matter one-halo term, so it does not depend on halo mass and from our suite of simulations we determine its Gaussian value to be $\simeq418h^{-3}\mathrm{Mpc}^3$ at $z=0$ and $\simeq45h^{-3}\mathrm{Mpc}^3$ at $z=1$. In the case of one single mass bin, Eq.~(\ref{E_hm}) reduces to $\mathcal{E}=\bar{n}^{-1}-2bM/\rhom+b^2\langle nM^2\rangle/\rhom^2$, while if we project out the lowest eigenmode $\Vm$ and normalize, we obtain the weighted shot noise
\begin{equation}
\mathcal{E}_w\equiv\frac{\Vm^\intercal\E\Vm\T}{\left(\Vm^\intercal\openone\right)^2}=\lambda_-\frac{\Vm^\intercal\Vm\T}{\left(\Vm^\intercal\openone\right)^2} \;.
\end{equation}
The eigenvalues $\lambda_\pm$ with eigenvectors $\Vpm$ can be found from Eq.~(\ref{E_hm}),
\begin{gather}
\lambda_\pm=\bar{n}^{-1}-\Mr^\intercal\bg\pm\sqrt{\Mr^\intercal\Mr\;\bg^\intercal\bg} \;, \\
\Vpm=\mathcal{N_\pm}^{-1}\left(\Mr\left/\sqrt{\Mr^\intercal\Mr}\right.\mp\bg\left/\sqrt{\bg^\intercal\bg}\right.\right) \;,
\end{gather}
where
\begin{equation}
\mathcal{N_\pm}\equiv\sqrt{2\mp2\Mr^\intercal\bg\left/\sqrt{\Mr^\intercal\Mr\;\bg^\intercal\bg}\right.} \;
\end{equation}
is a normalization constant to guarantee $\Vpm^\intercal\Vpm\T=1$. It is easily verified that $\Vpm^\intercal\Vmp\T=0$, i.e., they are orthogonal. In the continuous limit of infinitely many bins ($N\rightarrow\infty$) we can replace $\Vpm$ by the smooth function
\begin{equation}
V_{\pm}=\mathcal{N_\pm}^{-1}\left(\mathcal{M}\left/\sqrt{\langle\mathcal{M}^2\rangle}\right.\mp b\left/\sqrt{\langle b^2\rangle}\right.\right) \;,
\end{equation}
and obtain
\begin{gather}
\mathcal{E}_w=\left(\bar{n}_{\mathrm{tot}}^{-1}-\langle\mathcal{M}b\rangle-\sqrt{\langle\mathcal{M}^2\rangle\langle b^2\rangle}\right) \frac{\langle V_-^2\rangle}{\langle V_-\rangle^2} \;, \\
b_w=\frac{\langle V_-b\rangle}{\langle V_-\rangle}\;,
\end{gather}
where $b_w$ is the weighted effective bias and we exchanged the vector products by integrals over the mass function:
\begin{equation}
\boldsymbol{x}^\intercal\boldsymbol{y}\longrightarrow \frac{N}{\bar{n}_{\mathrm{tot}}}\int_{M_{\mathrm{min}}}^{M_\mathrm{max}}\frac{dn}{dM}(M)x(M)y(M)\;dM\equiv N\langle xy \rangle \;, \label{cont}
\end{equation}
\begin{equation}
{\bar{n}_{\mathrm{tot}}}=\int_{M_{\mathrm{min}}}^{M_\mathrm{max}}\frac{dn}{dM}(M)\;dM\equiv N\bar{n} \;.
\end{equation}

\begin{figure*}[!t]
\centering
\resizebox{\hsize}{!}{
\includegraphics{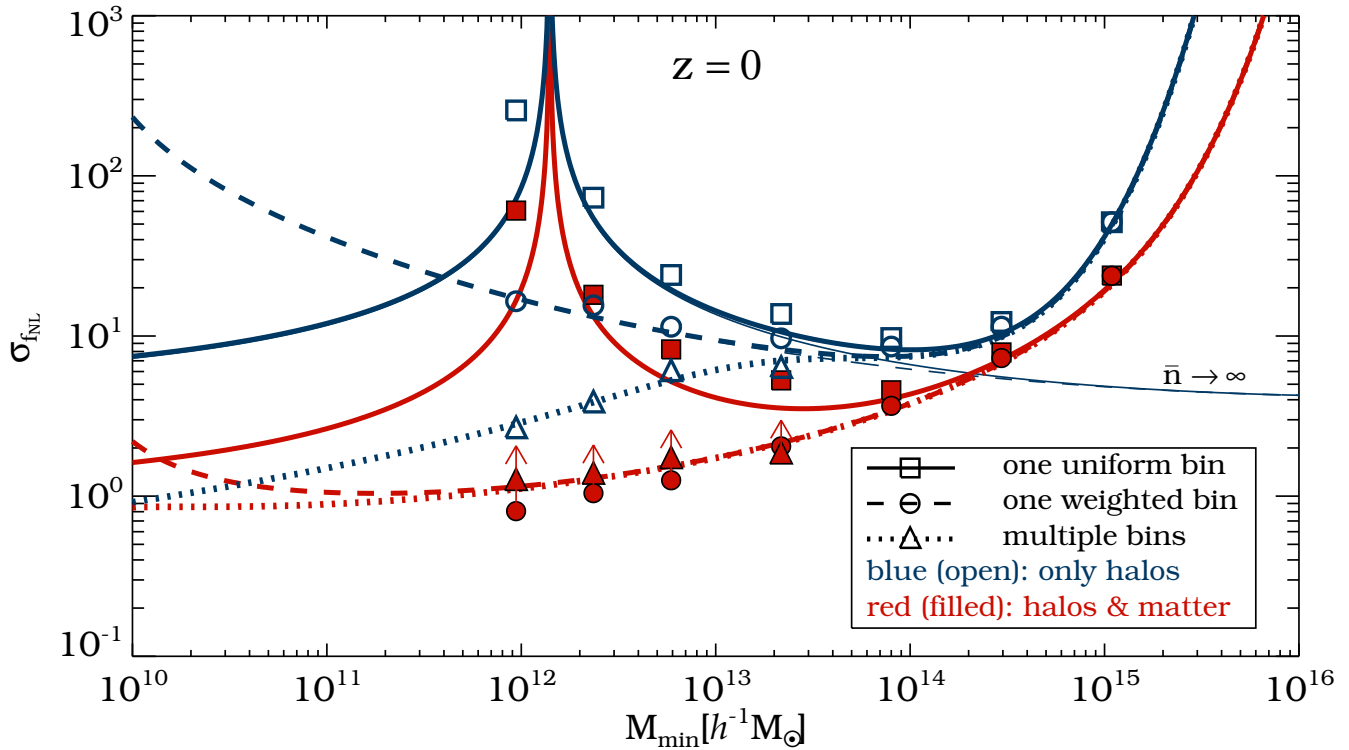}}
\caption{Halo model predictions for the one-sigma error on $\fnl$ (inferred from an effective volume of $V_{\mathrm{eff}}\simeq50h^{-3}\mathrm{Gpc}^3$, taking into account all modes with $0.0039h\mathrm{Mpc}^{-1}\le k\le0.032h\mathrm{Mpc}^{-1}$ at $z=0$) as a function of minimum halo mass from uniform- (solid lines) and weighted halos (dashed lines) from a single mass bin. The $N$-body simulation results are overplotted, respectively, as squares and circles for different low-mass cuts. Results that assume knowledge of halos and the dark matter are plotted in red (filled symbols), those that only consider halos are depicted in blue (open symbols). The dotted lines (triangles) show the results from splitting the halo catalog into multiple mass bins and taking into account the full halo covariance matrix in calculating $F_{\fnl\fnl}$. The high-sampling limit for one mass bin ($\bar{n}\rightarrow\infty$, $F_{\fnl\fnl}=2\left(b'/b\right)^2$) is overplotted for the uniform- (thin solid line) and the weighted case (thin dashed line). Arrows show the effect of adding a log-normal scatter of $\sigma_{\ln M}=0.5$ to all halo masses, they are omitted in all cases where the scatter has negligible impact.}
\label{HM_F_z=0}
\end{figure*}

Figure \ref{HM_b_sn} depicts the halo model prediction for the scale-independent Gaussian bias $b_{\mathrm{G}}$ and shot noise $\mathcal{E}$ as a function of minimum halo mass $M_{\mathrm{min}}$ at $z=0$ and $z=1$ for both the uniform and the weighted case of a single mass bin. Simulation results are overplotted as symbols for a few $M_{\mathrm{min}}$ [we approximate the weighting function $V_-(M)$ by $w(M)$ from Eq.~(\ref{w(M)}) in the simulations]. Obviously, modified mass weighting increases $b_{\mathrm{G}}$, especially when going to lower halo masses. It is also worth noticing that in contrast to the uniform case, $b_{\mathrm{G}}$ is always greater than unity when weighted by $w(M)$ (at least in the considered mass range). Going to higher redshift further increases $b_{\mathrm{G}}$ at any given $M_{\mathrm{min}}$.

For the shot noise we observe the opposite behavior: modified mass weighting leads to a suppression of~$\mathcal{E}$, which is increasingly pronounced towards lower halo masses. Moreover, it is always below the Poisson-model prediction of $\bar{n}_{\mathrm{tot}}^{-1}$. Our $N$-body simulation results generally confirm this trend (at least down to our resolution limit), although the halo model slightly underestimates the suppression of shot noise between uniform and weighted halos at lower $M_{\mathrm{min}}$. At higher redshifts, this suppression becomes smaller at given $M_{\mathrm{min}}$, but the magnitude of $\mathcal{E}_w$ at $z=1$ approaches the one at $z=0$ towards low $M_{\mathrm{min}}$ and is still small compared to the Poisson-model prediction of $\bar{n}_{\mathrm{tot}}^{-1}$.

\begin{figure*}[!t]
\centering
\resizebox{\hsize}{!}{
\includegraphics{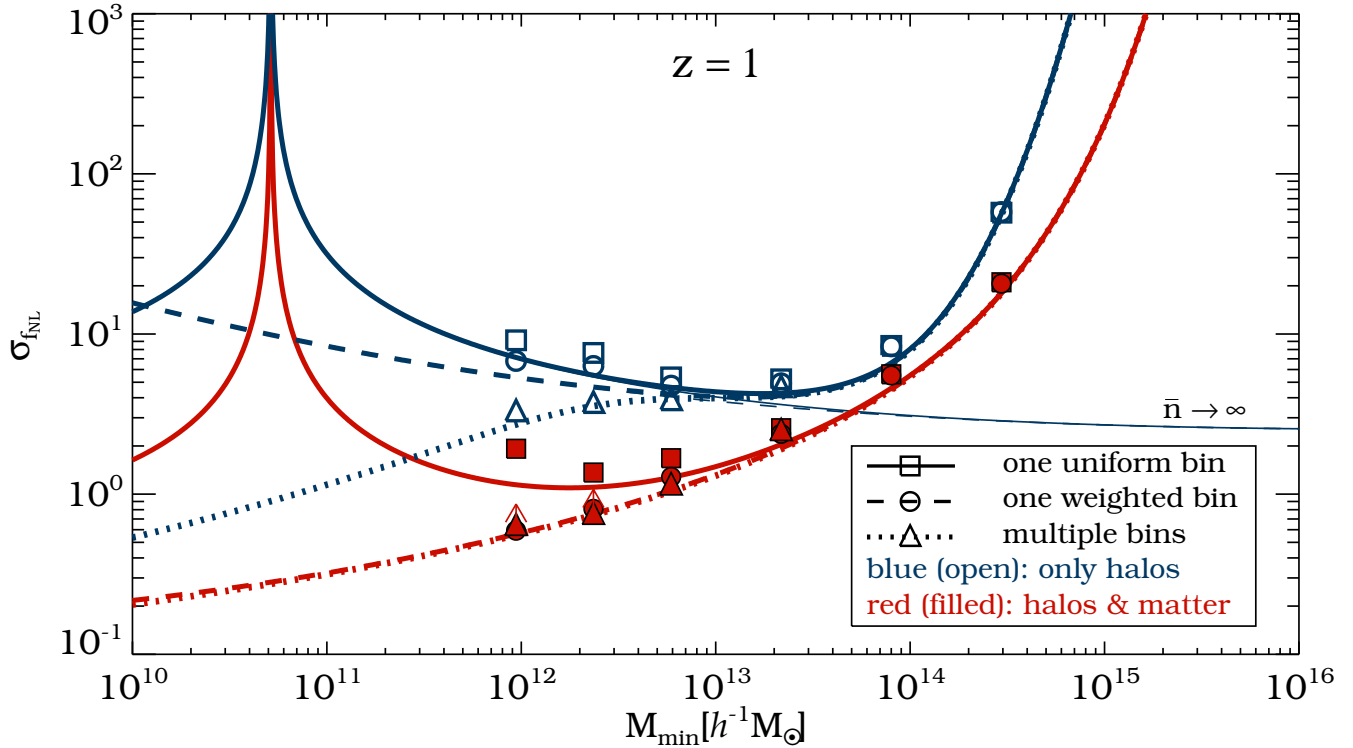}}
\caption{Same as Fig.~\ref{HM_F_z=0} at $z=1$.}
\label{HM_F_z=1}
\end{figure*}

\subsection{Single tracer}
With predictions for $b_{\mathrm{G}}$ and $\mathcal{E}$ at hand, we can directly compute the expected Fisher information content on $\fnl$ from a single halo mass bin. If the dark matter density field is known we apply Eq.~(\ref{F_m1}), if it is not we use Eq.~(\ref{F1_text}). In order to obtain most conservative results, we neglect terms featuring derivatives of the shot noise with respect to $\fnl$. We then apply Eq.~(\ref{sigma_fnl}) with $k_{\mathrm{min}}=0.0039h\mathrm{Mpc}^{-1}$, $k_{\mathrm{max}}=0.032h\mathrm{Mpc}^{-1}$ and $V\simeq50h^{-3}\mathrm{Gpc}^3$ to compute the one-sigma error forecast for $\fnl$.

Results are shown in Fig.~\ref{HM_F_z=0} for $z=0$. When the dark matter density field is available (red lines and filled symbols), weighting the halos (red dashed lines and filled circles) is always superior to the conventional uniform case (red solid lines and filled squares), especially when going to lower halo masses. In particular, $\sigma_{\fnl}$ substantially decreases with decreasing $M_{\mathrm{min}}$ in the weighted case, while for uniform halos it shows a spike at $M_{\mathrm{min}}\simeq1.4\times10^{12}\hMsun$. This happens when $b_{\mathrm{G}}$ becomes unity and the non-Gaussian correction to the halo bias in Eq.~(\ref{b(k,fnl)}) vanishes, leaving no signature of $\fnl$ in the effective bias. Since in the weighted case $b_{\mathrm{G}}>1$ for all considered $M_{\mathrm{min}}$, this spike does not appear, although we notice that the error on $\fnl$ begins to increase below $M_{\mathrm{min}}\sim10^{11}\hMsun$.

The simulation results are overplotted as symbols for a few values of $M_{\mathrm{min}}$, the agreement with the halo model predictions is remarkable. Note that the first two data-points at $M_{\mathrm{min}}=9.4\times10^{11}\hMsun$ and $M_{\mathrm{min}}=2.35\times10^{12}\hMsun$ resulting from our high-resolution simulation were scaled to the effective volume of our $12$ low-resolution boxes. The simulations yield a minimum error of $\sigma_{\fnl}\simeq0.8$ at $M_{\mathrm{min}}\simeq10^{12}\hMsun$ in the optimally weighted case with the dark matter available. This value is even lower than what is anticipated by the halo model ($\sigma_{\fnl}\simeq1$).

The results without the dark matter are shown as blue lines and open symbols. $\sigma_{\fnl}$ exhibits a minimum at $M_{\mathrm{min}}\simeq10^{14}\hMsun$ with $\sigma_{\fnl}\sim10$ for both uniform and weighted halos. Thus, weighting the halos does not decrease the lowest possible error on $\fnl$ from the uniform case, as expected. This suggests that only the highest-mass halos (clusters at $z=0$) need to be considered to optimally constrain $\fnl$ from a single-bin survey without observations of the dark matter.

In the limit of $\bar{n}\rightarrow\infty$, $F_{\fnl\fnl}\rightarrow2\left(b'/b\right)^2$. Then, according to Eq.~(\ref{b(k,fnl)}) for high $M_{\mathrm{min}}$, $b'\rightarrow b_{\mathrm{G}}u$, and hence $F_{\fnl\fnl}\rightarrow2u^2$ becomes independent of $M_{\mathrm{min}}$. The corresponding $\sigma_{\fnl}$ in the limit $\bar{n}\rightarrow\infty$ is plotted in Fig.~\ref{HM_F_z=0} for both uniform- (thin blue solid line) and weighted halos (thin blue dashed line) and it indeed approaches a constant value at high $M_{\mathrm{min}}$. It is about a factor of $2$ below the minimum in $\sigma_{\fnl}$ without setting $\bar{n}\rightarrow\infty$.

The results for redshift $z=1$ are presented in Fig.~\ref{HM_F_z=1}. In comparison to Fig.~\ref{HM_F_z=0} one can observe that all the curves are shifted towards the lower left of the plot, i.e., the constraints on $\fnl$ improve with increasing redshift. This is mainly due to the increase of the Gaussian effective bias $b_{\mathrm{G}}$ with $z$, as evident from the left panel of Fig.~\ref{HM_b_sn}. For example, the location of the spikes in $\sigma_{\fnl}(M_{\mathrm{min}})$ requires $b_{\mathrm{G}}=1$. At $z=1$ this condition is fulfilled at lower $M_{\mathrm{min}}$ ($\simeq5\times10^{10}\hMsun$) than at $z=0$, thus shifting the spikes to the left. Further, since the Fisher information from Eqs.~(\ref{F1_text}) and (\ref{F_m1}) increases with $b_{\mathrm{G}}$, $\sigma_{\fnl}(M_{\mathrm{min}})$ decreases, especially at low $M_{\mathrm{min}}$.

In the case of optimally weighted halos with knowledge of the dark matter, our simulations suggest $\sigma_{\fnl}\simeq0.6$ when reaching $M_{\mathrm{min}}\simeq 10^{12}\hMsun$ at $z=1$, in good agreement with the halo model. It even forecasts $\sigma_{\fnl}\simeq0.2$ when including halos down to $M_{\mathrm{min}}\simeq10^{10}\hMsun$.

\subsection{Multiple tracers}
The more general strategy for constraining $\fnl$ from a galaxy survey is to consider all auto- and cross-correlations between tracers of different mass, namely, the halo covariance matrix $\C$. If the dark matter density field is known, one can add the correlations with this field and determine $\Cm$. The Fisher information on $\fnl$ is then given by Eq.~(\ref{F_lin}) and Eq.~(\ref{F_m_text}), respectively. Again, the halo model can be applied to make predictions on the Fisher information content. In Appendix~\ref{appendix4}, the analytical expressions for $\alpha$, $\beta$ and $\gamma$ are derived for arbitrarily many mass bins and the continuous limit of infinite bins.

\begin{figure*}[!t]
\centering
\resizebox{\hsize}{!}{
\includegraphics[trim = 0 32 13 0,clip]{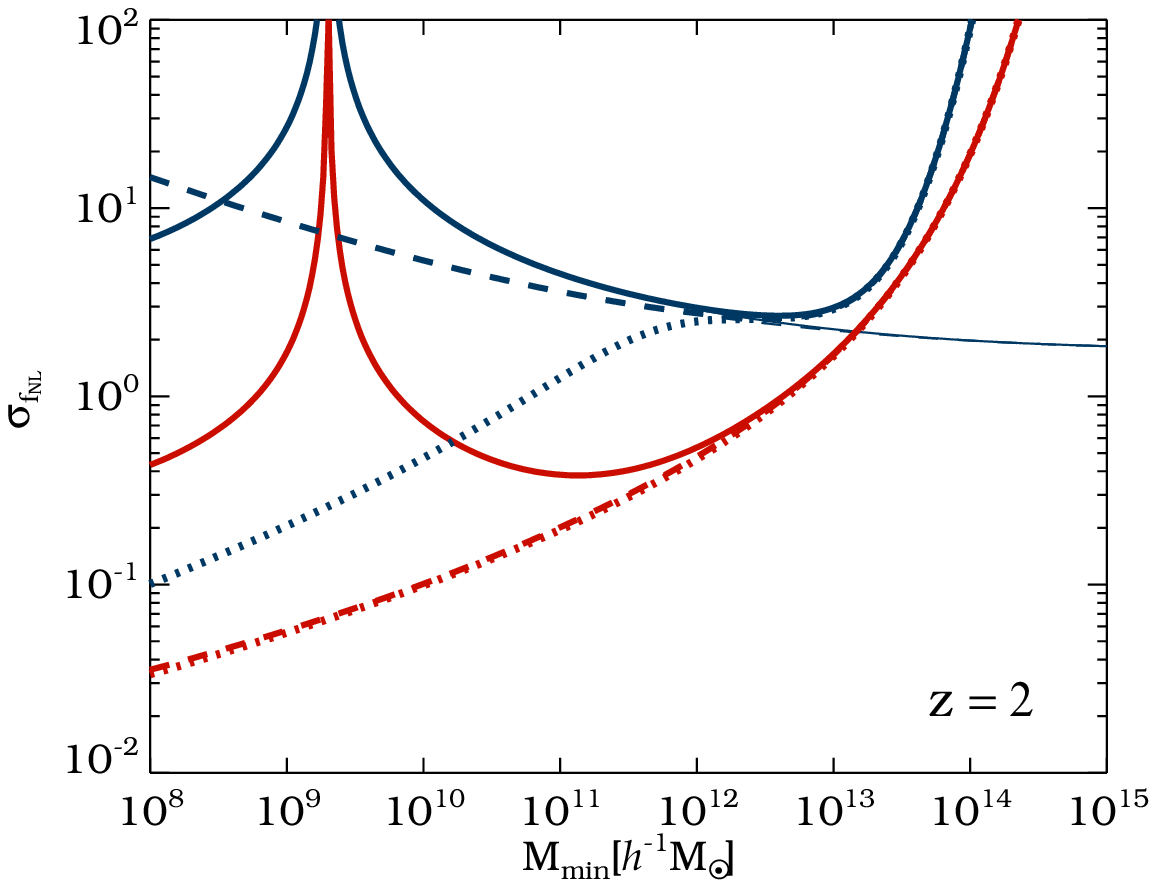}
\includegraphics[trim = 49 32 0 0,clip]{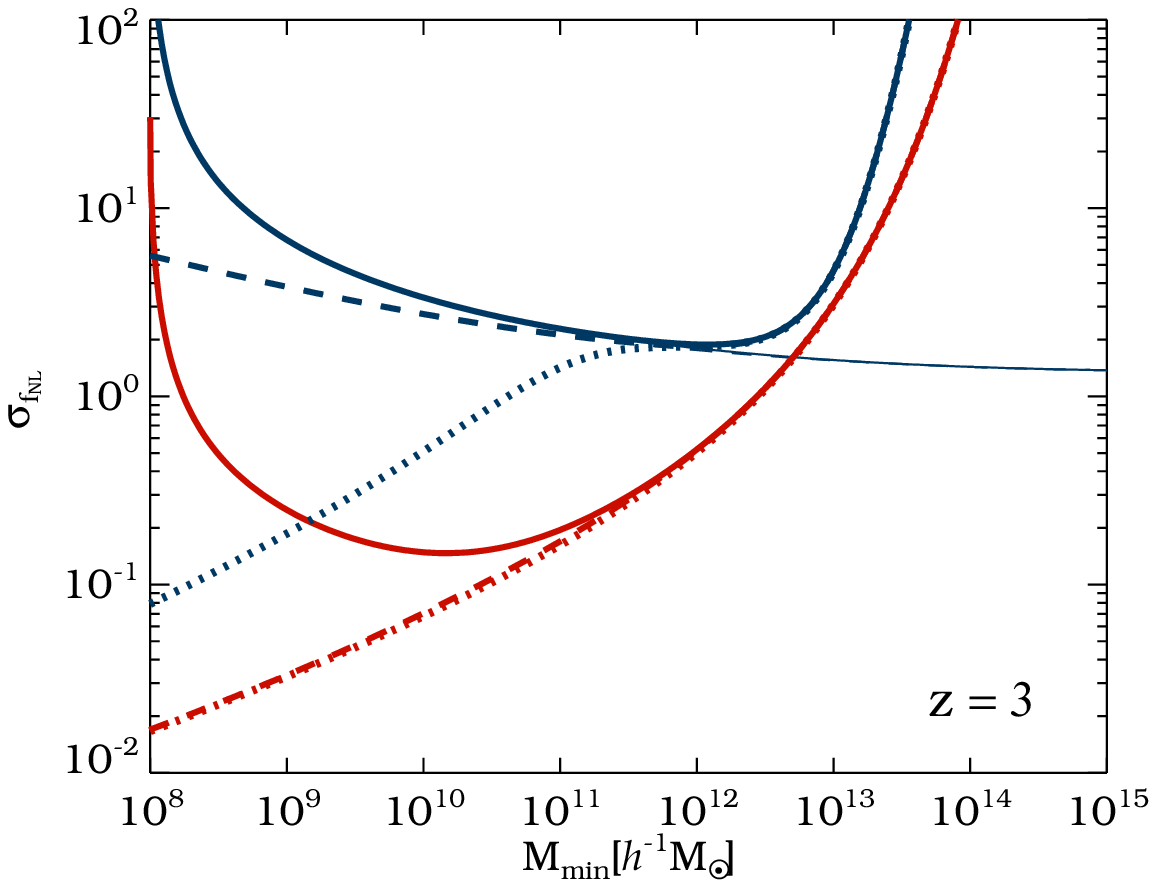}}
\resizebox{\hsize}{!}{
\includegraphics[trim = 0 0 13 2,clip]{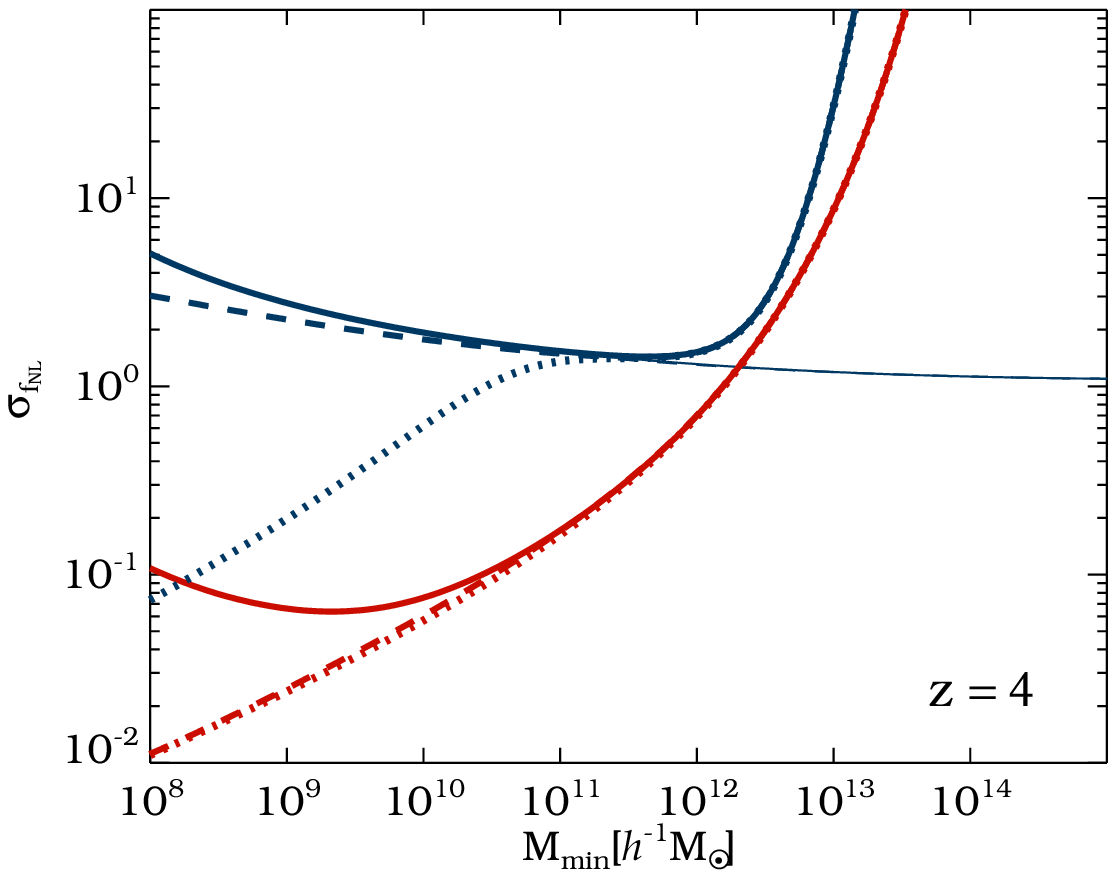}
\includegraphics[trim = 49 0 0 2,clip]{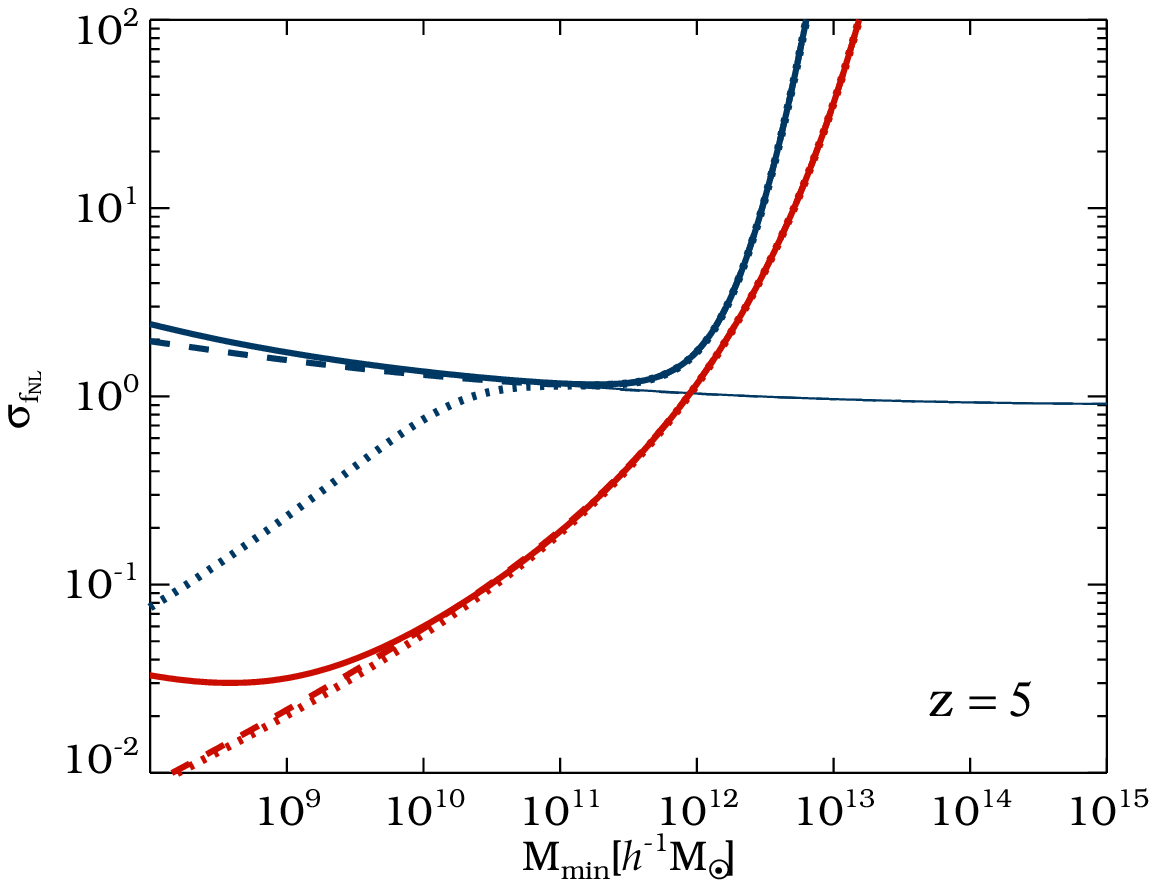}}
\caption{Same as Figs.~\ref{HM_F_z=0} and \ref{HM_F_z=1} at higher redshifts, as indicated in the bottom right of each panel. Here, only the halo model predictions are shown.}
\label{HM_F_z}
\end{figure*}

The dotted lines in Fig.~\ref{HM_F_z=0} show the halo model predictions at $z=0$ in this continuous limit of infinitely many mass bins. When the dark matter is available (red dotted line), $\sigma_{\fnl}$ coincides with the results from the optimally weighted one-bin case (dashed red lines). This confirms our claim that with the dark matter density field at hand, modified mass weighting is the optimal choice for constraining $\fnl$ and yields the maximal Fisher information content. Only below $M_{\mathrm{min}}\sim10^{12}\hMsun$ the optimally weighted halo field becomes slightly inferior to the case of infinite bins.

From multiple bins of halos without the dark matter (blue dotted line) we observe a different behavior. While at high $M_{\mathrm{min}}$ the error on $\fnl$ still matches the results from one mass bin, either uniform (blue solid line) or weighted (blue dashed line), below $M_{\mathrm{min}}\sim10^{14}\hMsun$ it departs towards lower values and finally reaches the same continuous limit as in the case where the dark matter is available at $M_{\mathrm{min}}\sim10^{10}\hMsun$. Thus, galaxies in principle suffice to yield optimal constraints on $\fnl$, however, one has to go to very low halo mass.

Our simulation results for multiple bins (triangles in Fig.~\ref{HM_F_z=0}) support this conclusion. Although we can only consider a limited number of mass bins in the numerical analysis (we used $N=30$ for our $12$ low-res boxes and $N=10$ for our high-res box), the continuous limit of the halo model can be approached closely. However, note that residuals of sampling variance in the numerical determination of $\E$, as described in Sec.~\ref{sec:halos&dm} and shown in Fig.~\ref{SN}, can result in an overestimation of $F_{\fnl\fnl}$. This is especially the case when the number of mass bins $N$ is high, resulting in a low halo number density per bin~$\bar{n}$. Hence, we chose $N$ such that the influence of sampling variance on our results is negligible, and yet clear improvements compared to the single-tracer case are established.

One concern in practical applications is scatter in the halo mass estimation. Although X-ray cluster-mass proxies show very tight correlations with halo mass with a log-normal scatter of $\sigma_{\ln M}\lesssim0.1$ \cite{Kravtsov2006,Fabjan2011}, optical mass-estimators are more likely to have $\sigma_{\ln M}\simeq0.5$ \cite{Rozo2009}. We applied a log-normal mass scatter of $\sigma_{\ln M}=0.5$ to all of our halo masses and repeated the numerical analysis for all the cases (symbols) shown in Fig.~\ref{HM_F_z=0}. The arrows in that figure show the effect of adding the scatter, pointing to the new (higher) value of $\sigma_{\fnl}$. We find the effect of the applied mass scatter to be negligible in most of the considered cases (arrows omitted). Only in the case of one weighted halo bin with knowledge of the dark matter (red filled circles) we observe a moderate weakening in $\fnl$-constraints, especially towards lower $M_{\mathrm{min}}$. This is expected, since we make most heavy use of the halo masses when applying modified mass weighting. Yet, the improvement compared to the uniform one-bin case remains substantial, so the method is still beneficial in the presence of mass scatter.

At higher redshifts, we observe the same characteristics as in the single-tracer case: the $\sigma_{\fnl}$-curves are shifted towards the lower left of the plot in Fig.~\ref{HM_F_z=1}, due to the increase in the effective bias with $z$. Moreover, the impact of mass scatter on $\sigma_{\fnl}$ becomes less severe at higher redshifts, as evident from the smaller arrows in Fig.~\ref{HM_F_z=1} as compared to Fig.~\ref{HM_F_z=0}. High-redshift data are therefore more promising for constraining $\fnl$. This is good news, since the relatively large effective volume assumed in the current analysis ($V_{\mathrm{eff}}\simeq50h^{-3}\mathrm{Gpc}^3$) can only be reached in practical applications when going to $z\sim1$ or higher.

On the other hand, the convergence of the constraints obtained with and without the dark matter is pushed to even lower halo masses at higher redshifts. This can be seen in Fig.~\ref{HM_F_z}, where we show the halo model predictions for even higher redshifts, going up to $z=5$. With a mass threshold of $M_{\mathrm{min}}=10^{10}\hMsun$, the optimal constraints on $\fnl$ from only halos start to saturate above $z\simeq2$, where $\sigma_{\fnl}\simeq0.5$. This is however not the case when the dark matter is available: the error on $\fnl$ decreases monotonically up to $z=5$ reaching $\sigma_{\fnl}\simeq0.06$, although for practical purposes it will be difficult to achieve this limit. Yet, reaching $\sigma_{\fnl}\sim1$ at $z=1$ and $M_{\mathrm{min}}\sim10^{11}\hMsun$ with a survey volume of about $50h^{-3}\mathrm{Gpc}^3$ seems realistic.

\section{Conclusions}
\label{sec:conclusion}
The aim of this work is to assess the amount of information on primordial non-Gaussianity that can be extracted from the two-point statistics of halo- and dark matter large-scale structure in light of shot noise suppression and sampling variance cancellation techniques that have been suggested in the literature. For this purpose we developed a theoretical framework for calculating the Fisher information content on $\fnl$ that relies on minimal assumptions for the covariance matrix of halos in Fourier space. The main ingredients of this model are the \emph{effective bias} and the \emph{shot noise matrix}, both of which we measure from $N$-body simulations and compare to analytic predictions. Our results can be summarized as follows:

\begin{itemize}
\item On large scales the effective bias agrees well with linear theory predictions from the literature, while towards smaller scales, it shows deviations that can be explained by the local bias-expansion model. The shot noise matrix exhibits
two nontrivial eigenvalues $\lambda_+$ and $\lambda_-$, both of which show a considerable dependence on $\fnl$. We further show that the eigenvector $\Vp$ is closely related to the second-order bias and that the corresponding eigenvalue $\lambda_+$ depends on the shot noise of the squared dark matter density field $\mathcal{E}_{\delta^2}$, which itself depends on $\fnl$ weakly. This property can become important when constraining $\fnl$ from very high-mass halos (clusters). However, since the Fourier modes of $\mathcal{E}_{\delta^2}$ are highly correlated, it is questionable how much information on primordial non-Gaussianity can be gained from the $\fnl$-dependence of the shot noise matrix. We demonstrate, though, that for the considered values of $\fnl$ the assumption of a Gaussian form of the likelihood function is sufficient to determine the correct Fisher information.

\item With the help of $N$-body simulations we demonstrate how the parameter $\fnl$ can be constrained and its error reduced relative to traditional methods by applying optimal weighting- and multiple-tracer techniques to the halos. For our specific simulation setup with $M_{\mathrm{min}}\sim10^{12}\hMsun$, we reach almost 1 order of magnitude improvements in $\fnl$-constraints at $z=0$, even if the dark matter density field is not available. The absolute constraints on $\fnl$ depend on the effective volume and the minimal halo mass that is resolved in the simulations, or observed in the data, and are expected to improve further when higher redshifts or lower-mass halos are considered.

\item We confirm the existence of a suppression factor (denoted $q$-factor in the literature) in the amplitude of the linear theory correction to the non-Gaussian halo bias. We argue that this only holds for halos generated with a friends-of-friends finding algorithm and depends on the specified linking length between halo particles. For a linking length of $20\%$ of the mean interparticle distance, our simulations yield $q\simeq0.8$. For halos generated with a spherical overdensity finder, we demonstrate that the best-fit values of $\fnl$ measured from the simulations are fairly consistent with the input values, i.e., $q\simeq 1$.

\item We calculate the Fisher information content from the two-point statistics of halos and dark matter in Fourier space, both analytically and numerically, and express the results in terms of an effective bias, a shot noise matrix and the dark matter power spectrum. In the case of a single mass bin and assuming knowledge of the dark matter density field, the Fisher information is inversely proportional to the shot noise and, therefore, not bounded from above if the shot noise vanishes. However, when only the halo distribution is available, the Fisher information remains finite even in the limit of zero shot noise. In this case, the amount of information on $\fnl$ can only be increased by dividing the halos into multiple mass bins (multiple tracers).

\item Utilizing the halo model we calculate $\sigma_{\fnl}$ and find a remarkable agreement with our simulation results. We show that in the continuous limit of infinite mass bins, optimal constraints on $\fnl$ can in principle be achieved even in the case where dark matter observations are not available. With an effective survey volume of $\simeq50h^{-3}\mathrm{Gpc}^3$ out to scales of $k_{\mathrm{min}}\simeq0.004h\mathrm{Mpc}^{-1}$ this means $\sigma_{\fnl}\sim1$ when halos down to $M_{\mathrm{min}}\sim10^{11}\hMsun$ are observed at $z=0$. In comparison to this, a single-tracer method yields $\fnl$-constraints that are weaker by about 1 order of magnitude. Further improvements are expected at higher redshifts and lower $M_{\mathrm{min}}$, potentially reaching the level of $\sigma_{\fnl}\lesssim0.1$.

\item  In realistic applications, additional sources of noise, such as a scatter in halo mass will have to be considered. We test the impact of adding a log-normal scatter of $\sigma_{\ln M}=0.5$ to our halo masses and find our results to be relatively unaffected. Assuming the dark matter to be available to correlate against halos is even more uncertain. Weak-lensing tomography can only measure the dark matter over a broad radial projection and more work is needed to see how far this approach can be pushed. Moreover, one would also need to include weak-lensing ellipticity noise into the analysis, which we have not done here.

\end{itemize}

We conclude that the shot noise suppression method (modified mass weighting) as presented in \cite{Hamaus2010} when the dark matter density field is available, and the sampling variance cancellation technique (multiple tracers) as proposed in \cite{Seljak2009a} when it is not, have the potential to significantly improve the constraints on primordial non-Gaussianity from current and future large-scale structure data. In \cite{Baldauf2011a} it was found (their Fig.~$15$) that while the power spectrum analysis of a single tracer with $M_{\mathrm{min}}\sim10^{14}\hMsun$ (close to our optimal mass for a single tracer without the dark matter) predicts $\sigma_{\fnl}\sim10$ for $V_{\mathrm{eff}}\simeq50h^{-3}\mathrm{Gpc}^3$, in good agreement with our results, the bispectrum analysis improves this to $\sigma_{\fnl}\sim5$. Our results suggest that the multitracer analysis of the halo power spectrum can improve upon a single-tracer bispectrum analysis, potentially reaching significantly smaller errors on $\fnl$. In principle the multitracer approach can also be applied to the halo bispectrum, but it is not clear how much one can benefit from it, since the dominant terms in the bispectrum do not feature any additional scale dependence that changes with tracer-mass.

In this paper we only focused on primordial non-Gaussianity of the local type and the two-point correlation analysis. Yet, our techniques can be applied to some, but not all, other models of primordial non-Gaussianity, which have only recently been studied in simulations \cite{Taruya2008,Bartolo2010a,Desjacques2010c,Wagner2010,Fedeli2011,Wagner2011}. Theoretical calculations of the non-Gaussian halo bias generally suggest different degrees of scale dependence and amplitudes depending on the model \cite{Verde2009,Sefusatti2009a,Schmidt2010,Shandera2011,Becker2011}. Our methods may help to test those various classes of primordial non-Gaussianity and thus provide a tool to probe the physics of the very early Universe.

\begin{acknowledgments}
We thank Pat McDonald, Tobias Baldauf, Ravi Sheth and Jaiyul Yoo for fruitful discussions, V. Springel for making public his N-body code {\scshape gadget ii}, and A. Knebe for making public his SO halo finder {\scshape AHF}. This work is supported by the Packard Foundation, the Swiss National Foundation under contract 200021-116696/1 and WCU grant R32-10130. VD acknowledges additional support from FK UZH 57184001. NH thanks the hospitality of Lawrence Berkeley National Laboratory (LBNL) at UC Berkeley and the Institute for the Early Universe (IEU) at Ewha University Seoul, where parts of this work were completed.
\end{acknowledgments}

\bibliography{ms.bib}
\bibliographystyle{h-physrev.bst}

\appendix

\onecolumngrid

\section{FISHER INFORMATION ON PRIMORDIAL NON-GAUSSIANITY FROM THE COVARIANCE OF HALOS \label{appendix1}}
\noindent
Plugging Eq.~(\ref{C_fnl}) into Eq.~(\ref{fisher}) and using the cyclicity of the trace yields
\begin{multline}
F_{\fnl\fnl}=\frac{1}{2}\mathrm{Tr}\left(\frac{\partial\C}{\partial\fnl}\C^{-1}\frac{\partial\C}{\partial\fnl}\C^{-1}\right)=\frac{1}{2}\mathrm{Tr}\left[\left(\bg'\bg^\intercal\C^{-1}P+\bg\bg'^\intercal\C^{-1}P+\E'\C^{-1}\right)^2\right]= \\
=\bg^\intercal\C^{-1}\bg\bg'^\intercal\C^{-1}\bg'P^2 + \bg^\intercal\C^{-1}\bg'\bg^\intercal\C^{-1}\bg'P^2 + 2\bg^\intercal\C^{-1}\E'\C^{-1}\bg' P + \frac{1}{2}\mathrm{Tr}\Big(\E'\C^{-1}\E'\C^{-1}\Big) \;.
\end{multline}
Applying Eq.~(\ref{Sherman-Morrison}) yields
\begin{gather}
\bg^\intercal\C^{-1}\bg P =
\alpha-\frac{\alpha^2}{1+\alpha} = \frac{\alpha}{1+\alpha} \;, \\
\bg^\intercal\C^{-1}\bg' P =
\beta-\frac{\beta\alpha}{1+\alpha} = \frac{\beta}{1+\alpha} \;, \\
\bg'^\intercal\C^{-1}\bg' P =
\gamma-\frac{\beta^2}{1+\alpha} = \frac{\gamma+\alpha\gamma-\beta^2}{1+\alpha} \;, \\
\bg^\intercal\C^{-1}\E'\C^{-1}\bg' P =
\nu-\frac{\alpha\nu+\beta\mu}{1+\alpha}+\frac{\alpha\beta\mu}{\left(1+\alpha\right)^2} \;, \\
\frac{1}{2}\mathrm{Tr}\Big(\E'\C^{-1}\E'\C^{-1}\Big) =
\tau+\frac{\rho}{1+\alpha}+\frac{\mu^2/2}{\left(1+\alpha\right)^2} \;,
\end{gather}
where
\begin{gather}
\alpha\equiv\bg^\intercal\E^{-1}\bg P \;,\; \beta\equiv\bg^\intercal\E^{-1}\bg' P \;,\; \gamma\equiv\bg'^\intercal\E^{-1}\bg' P \;, \\
\mu\equiv-\bg^\intercal\left(\E^{-1}\right)'\bg P \;,\; \nu\equiv-\bg^\intercal\left(\E^{-1}\right)'\bg' P \;, \\
\rho\equiv\bg^\intercal\left(\E^{-1}\right)'\E\left(\E^{-1}\right)'\bg P \;, \\
\tau\equiv\frac{1}{2}\mathrm{Tr}\left(\E'\E^{-1}\E'\E^{-1}\right) \;.
\end{gather}
Finally, we get
\begin{equation}
F_{\fnl\fnl}=\frac{\left(1+\alpha\right)\left(\alpha\gamma+2\nu-\rho\right)+\left(1-\alpha\right)\beta^2+\left(\mu/2-2\beta\right)\mu}{\left(1+\alpha\right)^2}+\tau \;. \label{F}
\end{equation}
For a single mass bin we have $\alpha\gamma=\beta^2$, $\alpha\nu=\beta\mu$, $\alpha\rho=\mu^2$, $\gamma\rho=\nu^2$ and $\alpha\tau=\rho/2$. In this case, Eq.~(\ref{F}) becomes
\begin{equation}
F_{\fnl\fnl} = 2\left(\frac{\beta+\sqrt{\tau/2}}{1+\alpha}\right)^2=2\left(\frac{bb'P+\mathcal{E}'/2}{b^2P+\mathcal{E}}\right)^2 \;. \label{F1}
\end{equation}

\section{FISHER INFORMATION ON PRIMORDIAL NON-GAUSSIANITY FROM THE COVARIANCE OF HALOS AND DARK MATTER \label{appendix2}}

\noindent
Now we need to work out Eq.~(\ref{fisher}) by plugging in Eq.~(\ref{Covm_fnl}) and (\ref{CovmI}). Let us first note that
\begin{equation}
\frac{\partial\Cm}{\partial\fnl}\Cm^{-1} =
\left( \begin{array}{cc}
-\beta & \bg'^\intercal\E^{-1}P \\
\bg'-\beta\bg-\E'\E^{-1}\bg\;\; & \;\;\bg\bg'^\intercal\E^{-1} P+\E'\E^{-1} \\
\end{array} \right)  \;.
\end{equation}
This yields
\begin{multline}
F_{\fnl\fnl}=\frac{1}{2}\mathrm{Tr}\left[\beta^2+\bg'^\intercal\E^{-1}\bg'P-\beta\bg'^\intercal\E^{-1}\bg P-\bg'^\intercal\E^{-1}\E'\E^{-1}\bg P+\bg'\bg'^\intercal\E^{-1}P-\beta\bg\bg'^\intercal\E^{-1}P\right. \\
\left.-\E'\E^{-1}\bg\bg'^\intercal\E^{-1}P+\;\bg\bg'^\intercal\E^{-1}\bg\bg'^\intercal\E^{-1}P^2+\bg\bg'^\intercal\E^{-1}\E'\E^{-1}P+\E'\E^{-1}\bg\bg'^\intercal\E^{-1}P+\E'\E^{-1}\E'\E^{-1}\right]= \\
=\frac{1}{2}\left(\beta^2+\gamma-\beta^2-\nu+\gamma-\beta^2-\nu+\beta^2+\nu+\nu+2\tau\right)=\gamma+\tau \;. \label{F_m}
\end{multline}

\section{CANCELLATION OF DARK MATTER DENSITY AND COSMIC VARIANCE \label{appendix3}}
\noindent
In the case where the dark matter density field is known, one can immediately see from the first term in the chi-square of Eq.~(\ref{chi2m}), that with the model $\dhalo = \bg\dm+\e$ from Eq.~(\ref{dhalo}), the underlying density field $\delta$ is completely canceled (including its sampling variance) and the residual is
\begin{equation}
\langle\chi^2\rangle=\langle\e^\intercal\E^{-1}\e\rangle=\mathrm{Tr}\left(\E^{-1}\langle\e\e^\intercal\rangle\right)=N \;, \label{residual}
\end{equation}
where $N$ is the number of halo bins. If we only consider halos, the first term from Eq.~(\ref{chi2}) reads
\begin{equation}
\chi^2=\dhalo^\intercal\C^{-1}\dhalo\T=\dhalo^\intercal\E^{-1}\dhalo\T-\frac{P}{1+\alpha}\left(\dhalo^\intercal\E^{-1}\bg\right)^2 \;,
\end{equation}
where we used Eq.~(\ref{Sherman-Morrison}) in the second equality. Plugging in the model $\dhalo = \bg\dm+\e$, we get
\begin{multline}
\chi^2=\bg^\intercal\E^{-1}\bg\delta^2+2\bg^\intercal\E^{-1}\e\delta+\e^\intercal\E^{-1}\e-\frac{P}{1+\alpha}\left(\bg^\intercal\E^{-1}\bg\delta+\bg^\intercal\E^{-1}\e\right)^2= \\
\left(\alpha-\frac{\alpha^2}{1+\alpha}\right)\frac{\delta^2}{P}+2\bg^\intercal\E^{-1}\e\left(1-\frac{\alpha}{1+\alpha}\right)\delta+\e^\intercal\E^{-1}\e-\frac{P}{1+\alpha}\left(\bg^\intercal\E^{-1}\e\right)^2 \;. \label{residual_h}
\end{multline}
A large fraction of the first two terms in the last expression obviously cancel when $\alpha\gg1$. The quantity $\alpha$, also denoted as \emph{signal-to-noise ratio} in \cite{Hamaus2010}, monotonically increases with the number of halo bins $N$. In \cite{Hamaus2010} it was shown to reach $\mathcal{O}(10^2)$ in the continuous limit. Even higher values can be reached when the mass resolution of the simulation is increased \cite{Cai2011}. Hence, in the limit $\alpha\gg1$ the residual of the chi-square in Eq.~(\ref{residual_h}) becomes
\begin{equation}
\langle\chi^2\rangle=\frac{\langle\delta^2\rangle}{P}+2\bg^\intercal\E^{-1}\e\;\frac{\langle\delta\rangle}{\alpha}+\langle\e^\intercal\E^{-1}\e\rangle-\frac{P}{\alpha}\langle\left(\bg^\intercal\E^{-1}\e\right)^2\rangle=1+N-\frac{\bg^\intercal\E^{-1}\langle\e\e^\intercal\rangle\E^{-1}\bg}{\bg^\intercal\E^{-1}\bg}=N \;,
\end{equation}
the same as in Eq.~(\ref{residual}) with knowledge of the dark matter. Note that in the case of one single halo bin as in Eq.~(\ref{chi2_1}), a cancellation neither of the underlying dark matter field, nor of the sampling variance is possible.

\section{HALO MODEL PREDICTION FOR ALPHA, BETA AND GAMMA \label{appendix4}}
\noindent
In the halo model the shot noise matrix is given by Eq.~(\ref{E_hm}). In order to invert $\E$, we write $\E=\Z-\Mr\bg^\intercal$ with $\Z\equiv\bar{n}^{-1}\I-\bg\Mr^\intercal$ and apply the Sherman-Morrison formula:
\begin{equation}
\E^{-1}=\Z^{-1}+\frac{\Z^{-1}\Mr\bg^\intercal\Z^{-1}}{1-\bg^\intercal\Z^{-1}\Mr} \;.
\end{equation}
Likewise, we apply the Sherman-Morrison formula to invert $\Z$:
\begin{equation}
\Z^{-1}=\bar{n}\I+\frac{\bg\Mr^\intercal\bar{n}}{\bar{n}^{-1}-\Mr^\intercal\bg} \;.
\end{equation}
With
\begin{gather}
\alpha\equiv\bg^\intercal\E^{-1}\bg P=\frac{\bg^\intercal\Z^{-1}\bg}{1-\bg^\intercal\Z^{-1}\Mr}P \;, \\
\beta\equiv\bg^\intercal\E^{-1}\bg'P=\frac{\bg^\intercal\Z^{-1}\bg'}{1-\bg^\intercal\Z^{-1}\Mr}P \;, \\
\gamma\equiv\bg'^\intercal\E^{-1}\bg'P=\frac{\bg'^\intercal\Z^{-1}\bg'\left(1-\bg^\intercal\Z^{-1}\Mr\right)+\bg'^\intercal\Z^{-1}\Mr\bg^\intercal\Z^{-1}\bg'}{1-\bg^\intercal\Z^{-1}\Mr}P \;,
\end{gather}
and
\begin{gather}
\bg^\intercal\Z^{-1}\bg=\frac{\bg^\intercal\bg}{\bar{n}^{-1}-\Mr^\intercal\bg} \;, \\
\bg^\intercal\Z^{-1}\Mr=\frac{\Mr^\intercal\bg\left(\bar{n}^{-1}-\Mr^\intercal\bg\right)+\bg^\intercal\bg\Mr^\intercal\Mr}{\bar{n}^{-1}-\Mr^\intercal\bg}\bar{n} \;, \\
\bg^\intercal\Z^{-1}\bg'=\frac{\bg^\intercal\bg'\left(\bar{n}^{-1}-\Mr^\intercal\bg\right)+\bg^\intercal\bg\Mr^\intercal\bg'}{\bar{n}^{-1}-\Mr^\intercal\bg}\bar{n} \;, \\
\bg'^\intercal\Z^{-1}\Mr=\frac{\Mr^\intercal\bg'\left(\bar{n}^{-1}-\Mr^\intercal\bg\right)+\bg^\intercal\bg'\Mr^\intercal\Mr}{\bar{n}^{-1}-\Mr^\intercal\bg}\bar{n} \;, \\
\bg'^\intercal\Z^{-1}\bg'=\frac{\bg'^\intercal\bg'\left(\bar{n}^{-1}-\Mr^\intercal\bg\right)+\bg^\intercal\bg'\Mr^\intercal\bg'}{\bar{n}^{-1}-\Mr^\intercal\bg}\bar{n} \;,
\end{gather}
after some algebra we get
\begin{gather}
\alpha=\frac{\bg^\intercal\bg}{\lambda_+\lambda_-}\bar{n}^{-1}P \;, \\
\beta=\frac{\bg^\intercal\bg'\left(\bar{n}^{-1}-\Mr^\intercal\bg\right)+\bg^\intercal\bg\Mr^\intercal\bg'}{\lambda_+\lambda_-}P \;, \\
\gamma=\bg'^\intercal\bg'\bar{n}P+\frac{\bg^\intercal\bg\left(\Mr^\intercal\bg'\right)^2+\Mr^\intercal\Mr\left(\bg^\intercal\bg'\right)^2+2\bg^\intercal\bg'\Mr^\intercal\bg'\left(\bar{n}^{-1}-\Mr^\intercal\bg\right)}{\lambda_+\lambda_-}\bar{n}P \;,
\end{gather}
with $\lambda_+\lambda_-=\left(\bar{n}^{-1}-\Mr^\intercal\bg\right)^2-\bg^\intercal\bg\Mr^\intercal\Mr$. According to Eq.~(\ref{b(k,fnl)}) we can write $\bg'=\left(\bg-\openone\right)u$. Moreover, in the continuous limit we can exchange the vector products by integrals as in Eq.~(\ref{cont}). This finally yields
\begin{equation}
\alpha=\frac{\langle b^2\rangle}{\left(\bar{n}_{\mathrm{tot}}^{-1}-\langle\mathcal{M}b\rangle\right)^2-\langle b^2\rangle\langle\mathcal{M}^2\rangle}\bar{n}_{\mathrm{tot}}^{-1}P \;,
\end{equation}
\begin{equation}
\beta=\frac{\left(\langle b^2\rangle-\langle b\rangle\right)\left(\bar{n}_{\mathrm{tot}}^{-1}-\langle\mathcal{M}b\rangle\right)+\langle b^2\rangle\left(\langle\mathcal{M}b\rangle-\langle\mathcal{M}\rangle\right)}{\left(\bar{n}_{\mathrm{tot}}^{-1}-\langle\mathcal{M}b\rangle\right)^2-\langle b^2\rangle\langle\mathcal{M}^2\rangle}uP \;,
\end{equation}
\begin{align}
\gamma&=\left(\langle b^2\rangle-2\langle b\rangle+1\right)\bar{n}_{\mathrm{tot}}u^2P \nonumber \\
&\quad+ \frac{\langle b^2\rangle\left(\langle\mathcal{M}b\rangle-\langle\mathcal{M}\rangle\right)^2+\langle\mathcal{M}^2\rangle\left(\langle b^2\rangle-\langle b\rangle\right)^2+2\left(\langle b^2\rangle-\langle b\rangle\right)\left(\langle\mathcal{M}b\rangle-\langle\mathcal{M}\rangle\right)\left(\bar{n}_{\mathrm{tot}}^{-1}-\langle\mathcal{M}b\rangle\right)}{\left(\bar{n}_{\mathrm{tot}}^{-1}-\langle\mathcal{M}b\rangle\right)^2-\langle b^2\rangle\langle\mathcal{M}^2\rangle}\bar{n}_{\mathrm{tot}}u^2P \;.
\end{align}

\end{document}